\documentstyle[pra,aps,cite]{revtex}

\newcommand{\imina}{} 
\newcommand{\iminb}{} 
\newcommand{\ia}{} 
\newcommand{\ib}{} 
\begin{document}

\title{Excited states of the hydrogen molecule in magnetic fields:\\
The singlet $\Sigma$ states of the parallel configuration}
\author{P. Schmelcher, T.Detmer and L. S. Cederbaum}
\address{Theoretische Chemie, Physikalisch--Chemisches Institut,\\
Universit\"at Heidelberg, INF 229, D--69120 Heidelberg,\\
Federal Republic of Germany}
\maketitle

\begin{abstract}
Excited states of the hydrogen molecule subject to a homogeneous magnetic
field are investigated for the parallel configuration in the complete regime
of field strengths $B=0-100a.u.$. Up to seven excitations are studied for gerade as well
as ungerade spin singlet states of $\Sigma$ symmetry with a high accuracy.
The evolution of the potential energy curves for the individual states with 
increasing field strength as well as the overall behaviour of the spectrum are discussed in detail.
A variety of phenomena like for example the sequence of changes for the dissociation channels of
excited states and the resulting formation of outer wells are encountered.
Possible applications of the obtained data to the analysis of magnetic white dwarfs
are outlined.

\end{abstract}
\pacs{xxxx}

\section{Introduction} 

Matter which is exposed to strong external magnetic fields
changes its basic properties and structure
and leads to a variety of new phenomena. As a result strong fields are of importance in different
branches of physics like atomic, molecular or solid state physics. 
For atomic and molecular systems there are two prominent possibilities to encounter the strong field regime:
highly excited Rydberg states in the laboratory and atoms and molecules in the atmospheres of magnetized white dwarfs
(see refs.\cite{Friedrich:1989,Ruder:1994,Schmelcher:1998} for a compilation of the subject).  
From a theorists point of view particle systems in strong fields pose a hard problem
due to several competing interactions (Coulomb attraction and repulsion, para- and
diamagnetic interactions). Of particular interest, but most complicated to investigate, is hereby the 
so-called intermediate regime which is characterized by comparable magnetic and Coulomb binding
forces. Focusing on the low-lying states of atoms and molecules we envisage this regime 
for those magnetic white dwarfs which possess field strengths in the regime $10^{3}-10^{5}T$.
Each magnetic white dwarf possesses a characteristic regime of field strengths which varies, in case of a dipole,
by a factor of two from the pole to the equator. 
To perform a first identification of observed spectra from the atmospheres of these objects one uses the so-called
stationary line spectroscopy: characteristic absorption features can appear only for those wavelengths
which correspond to an extremum of the transition wavelength with respect to the field strength. 
In a second step one then performs simulations of the radiation transport in the atmosphere in order
to obtain synthetic spectra. In the eighties the above approaches have been used to identify hydrogen in a number
of magnetized white dwarfs \cite{Forster:1984,Henry:1984,Wickramasinghe:1988,Jordan:1989}.
Recently the stationary line argument has been successfully \cite{Jordan:1998} used to
obtain strong evidence for helium in the spectrum of the magnetic white dwarf GD229 whose absorption features have been
mysterious ever since its discovery $25$ years ago. This was only possible due to the enormous progress
achieved with respect to our knowledge of the spectrum and transitions of the helium atom in strong magnetic fields
\cite{Becken:1999,Becken:2000}.
However, this should not obscur the fact that there are a number of magnetic white dwarfs whose spectra
remain unexplained and furthermore new magnetic objects are discovered (see, for example, Reimers et al \cite{Reimers:1998}
in the course of the Hamburg survey of the European Southern Observatory (ESO)). Very recently strong candidates
for quasimolecular absorption features have been discovered in magnetized white dwarfs \cite{Jordan:1999}.
This raises the demand for a theoretical investigation of molecular properties, in particular of the
hydrogen molecule, in such strong fields.

Several theoretical investigations were performed for molecular systems in strong magnetic fields.
Most of them deal with the electronic structure of the $H_2^+$ ion (see refs. 
\cite{wille:1988,kappes1:1996,kappes2:1996,kappes3:1996,kravchenko:1997,kappes1:1994,kappes:1995}
and references therein). Very interesting phenomena can be observed 
already for this simple diatomic system. For the ground state of the $H_2^+$ molecule
the dissociation energy increases and the equilibrium internuclear distance simultaneously 
decreases with increasing field strength.
Furthermore it was shown \cite{kappes1:1994,kappes:1995} that a certain class of excited
electronic states, which possess a purely repulsive potential energy surface
in the absence of a magnetic field, acquire a well-pronounced potential well
in a sufficiently strong magnetic field.
Moreover the electronic potential energies depend not only on the internuclear distance
but also on the angle between the magnetic field and the molecular axis which leads to a very
complex topological behavior of the corresponding potential energy surfaces
\cite{kappes1:1996,kappes2:1996,kappes3:1996}.

In contrast to the $H_2^+$ ion there exist only a few investigations dealing with
the electronic structure of the hydrogen molecule in the presence of a strong magnetic field.
Highly excited states of $H_2$ were studied for a field strength of $4.7\;T$ in ref.
\cite{monteiro:1990}. For intermediate field strengths two studies of almost qualitative character
investigate the potential energy curve (PEC) of the lowest $^1\Sigma_g$ state
\cite{basile:1987,turbiner:1983}. A few investigations were performed in the high field limit
\cite{korolev:1992,lai:1992,ortiz:1995,lai:1996}, where the magnetic forces
dominate over the Coulomb forces and therefore several approximations can be used.
Very recently a first step has been done in order to elucidate the electronic structure
of the $H_2$ molecule for the parallel configuration, i.e. for parallel internuclear and
magnetic field axes \cite{detmer1:1997,detmer1:1998,kravchenko1:1997,kravchenko1:1998}.
Hereby refs.\cite{detmer1:1997,detmer1:1998} apply an exact, i.e. fully correlated 
approach, whereas refs.\cite{kravchenko1:1997,kravchenko1:1998} use Hartree-Fock calculations 
and focus exclusively on the identification of the global ground state of the molecule.
In refs.\cite{detmer1:1997,detmer1:1998} the lowest states of the
$\Sigma$ and $\Pi$ manifolds were studied for gerade and ungerade parity as well as singlet and triplet spin symmetry.
Hereby accurate adiabatic electronic energies were obtained for a broad range of field strengths from field free space up to
strong magnetic fields of $100\;a.u.$
A variety of interesting effects were revealed. As in the case of the $H_2^+$ ion,
the lowest strongly bound states of $\Sigma$ symmetry,
i.e. the lowest $^1\Sigma_g$ , $^3\Sigma_g$  and $^1\Sigma_u$  state, show a decrease of the
bond length and an increase of the dissociation energy for sufficiently
strong fields. Furthermore a change in the dissociation channel occurs for the lowest $^1\Sigma_u$  state
between $B = 10.0$ and $20.0\;a.u.$ due to the existence of strongly bound $H^-$ states
in the presence of a magnetic field. The $^3\Sigma_g$  state was shown to exhibit an additional
outer minimum for intermediate field strengths which could provide vibrationally bound states.

An important result of refs. \cite{detmer1:1997,kravchenko1:1997} is the change of the ground state from the
lowest $^1\Sigma_g$ state to the lowest $^3\Sigma_u$ state between $B = 0.1$ and $0.2\;a.u.$
This crossing is of particular relevance for the binding properties of the global ground state of the molecule:
The $^3\Sigma_u$  state is an unbound state and possesses only a very shallow
van der Waals minimum which does not support any vibrational level.
Therefore, the global ground state of the hydrogen molecule for the parallel configuration is an
unbound state for $B \gtrsim 0.2\;a.u.$. Furthermore it has been shown
in ref. \cite{ortiz:1995} that for very strong fields
($B \gtrsim 3\times10^3\;a.u.$) the strongly bound $^3\Pi_u$ state is the global ground state of
the hydrogen molecule oriented parallel to the magnetic field.
Finally the complete scenario for the crossovers of the global ground state of the parallel configuration has been clarified
in ref.\cite{detmer1:1998} which contains the transition field strengths for the crossings among
the lowest states of $^1\Sigma_g,^3\Sigma_u$ and $^3\Pi_u$ symmetry.

The above considerations show that detailed studies of the electronic
properties of the hydrogen molecule in a magnetic field are very desirable.
The present investigation deals with the excited $\Sigma$ states of the hydrogen molecule in the 
parallel configuration which is distinct by its higher symmetry
\cite{schmelcher:1990}. We employ a full configuration interaction (CI) approach which is most
suitable for obtaining detailed information on the electronic structure.
Our investigation is divided into two separate studies: the first and present work focuses on singlet states
of both gerade and ungerade symmetry whereas a later investigation will focus on the corresponding triplet states again
for both parities. Due to the spin Zeeman splitting the spin singlet and triplet manifolds
are increasingly separated with increasing field strength. The spin character provides therefore a natural
dividing line of our extensive work which contains a large amount of information and data
on the behaviour of the excited states of the molecule. 
The results of our calculations include accurate adiabatic PECs for the complete
range of field strengths $0 \leq B \leq 100\;a.u.$. Up to seven excited
states have been investigated for each symmetry. We present detailed data
for the total and dissociation energies at the equilibrium internuclear distances as well as
the equilibrium positions themselves for the
lowest three excitations for each symmetry.
Due to the large amount of data the evolution of the higher excited states with
increasing field strength is presented only graphically. However, further informations like, for example,
the positions of the maxima and the accurate heights of the barriers or the complete data
of the PEC's as well as numerical data on the higher excited states, more precisely, on
the fourth up to the seventh excited state, can be obtained from the authors upon request.

In detail the paper is organized as follows. In section II  we describe the theoretical aspects
of the present investigation, including a discussion of the Hamiltonian, a description
of the atomic orbital basis set and some remarks on the CI approach. Section III contains
the results and an elaborate discussion of the evolution of the electronic structure in 
the presence of the magnetic field with increasing strength.

\section{Theoretical aspects} 

Our starting point is the total nonrelativistic molecular Hamiltonian in Cartesian coordinates.
The total pseudomomentum is a constant of motion and therefore commutes with the Hamiltonian
\cite{johnson:1983,avron:1978}.
For that reason the Hamiltonian can be simplified by performing a
so-called pseudoseparation of the center of mass motion
\cite{schmelcher2:1988,schmelcher:1994,johnson:1983} which introduces
the center of mass coordinate and the conserved
pseudomomentum as a pair of canonical conjugated variables.
Further simplifications can be achieved by a consecutive series of unitary transformations
\cite{schmelcher2:1988,schmelcher:1994}.

In order to separate the electronic and nuclear motion we perform the
Born-Oppenheimer approximation in the presence of a magnetic field
\cite{schmelcher1:1988,schmelcher2:1988,schmelcher:1994}.
As a first order approximation we assume infinitely heavy masses for the nuclei.
The origin of our coordinate system coincides with the midpoint of
the internuclear axis of the hydrogen molecule
and the protons are located on the $z$ axis.
The magnetic field is chosen parallel to the $z$ axis of our coordinate system and
the symmetric gauge is adopted for the vector potential.
The gyromagnetic factor of the electron is chosen to be equal to two.
The Hamiltonian, therefore, takes on the following appearance:
\begin{equation}
H =\sum\limits_{i=1}^2 \left\{\frac{1}{2}\bbox{p}_i^2
+ \frac{1}{8}\left( \bbox{B}\times \bbox{r}_i\right)^2
+ \frac{1}{2}\bbox{L}_i \bbox{B}- \frac{1}{|\bbox{r}_i - \bbox{R}/2|}
- \frac{1}{|\bbox{r}_i + \bbox{R}/2|} \right\} \\
+ \frac{1}{|\bbox{r}_1 - \bbox{r}_2|} + \frac{1}{R}
+ \bbox{S} \bbox{B}\label{form1}
\end{equation}
The symbols $\bbox{r}_i$, $\bbox{p}_i$ and $\bbox{L}_i$ denote the position vectors, their
canonical conjugated momenta and the angular momenta of the two electrons, respectively.
$\bbox{B}$ and $\bbox{R}$ are the vectors of the magnetic field and internuclear distance,
respectively and $R$ denotes the magnitude of $\bbox{R}$.
With $\bbox{S}$ we denote the vector of the total electronic spin.
Throughout the paper we will use atomic units.

The Hamiltonian (\ref{form1}) commutes with the following independent operators:
the parity operator $P$,
the projection $L_z$ of the electronic angular momentum on the internuclear axis,
the square $S^2$ of the total electronic spin and
the projection $S_z$ of the total electronic spin on the internuclear axis.
In field free space we encounter an additional independent symmetry namely the reflections
of the electronic coordinates at the $xz$ ($\sigma_v$) plane.
The eigenfunctions possess the corresponding eigenvalues $\pm1$.
This symmetry does not hold in the presence of a magnetic field!
Therefore, the resulting symmetry groups for the hydrogen molecule are $D_{\infty h}$ in 
field free space and $C_{\infty h}$ in the presence of a magnetic field \cite{schmelcher:1990}.

In order to solve the fixed-nuclei electronic Schr\"odinger equation
belonging to the Hamiltonian (\ref{form1}) we expand the electronic
eigenfunctions in terms of molecular configurations.
In a first step the total electronic eigenfunction $\Psi_{tot}$ of the Hamiltonian (\ref{form1})
is written as a product of its spatial part $\Psi$ and its spin part
$\chi$, i.e. we have $\Psi_{tot} = \Psi \chi$.
For the spatial part $\Psi$ of the wave function we use the LCAO-MO-ansatz, i.e.
we decompose $\Psi$ with respect to molecular orbital configurations $\psi$ of $H_2$,
which respect the corresponding symmetries (see above) and the Pauli principle:
\begin{eqnarray}
\Psi &=& \sum\limits_{i,j} c_{ij} \left[\psi_{ij}\left(\bbox{r}_1,\bbox{r}_2\right) \pm
\psi_{ij}\left(\bbox{r}_2,\bbox{r}_1\right)\right] \nonumber \\
&=&
\sum\limits_{i,j} c_{ij} \left[\Phi_i\left(\bbox{r_1}\right)\Phi_j\left(\bbox{r_2}\right)
\pm \Phi_i\left(\bbox{r_2}\right)\Phi_j\left(\bbox{r_1}\right)\right] \nonumber
\end{eqnarray}
The molecular orbital configurations $\psi_{ij}$ of $H_2$ are products of the corresponding
one-electron $H_{2}^+$ molecular orbitals $\Phi_i$ and $\Phi_j$.
The $H_{2}^+$ molecular orbitals are built from atomic orbitals
centered at each nucleus. A key ingredient of this procedure is a
basis set of nonorthogonal optimized
nonspherical Gaussian atomic orbitals which has been established previously
\cite{schmelcher3:1988,kappes2:1994}.
For the case of a $H_2-$molecule parallel to the magnetic field
these basis functions read as follows:

\begin{equation}
\phi^m_{kl} \left(\rho,z,\alpha,\beta,\pm R/2 \right) =
\rho^{|m|+2k} \left(z\mp R/2\right)^l exp\left\{-\alpha \rho^2 - \beta\left( z\mp R/2 \right)^2 \right\}
exp\left\{im\phi\right\} \label{form2}
\end{equation}

The symbols $\rho = + \sqrt{x^2+y^2}$ and $z$ denote the electronic coordinates.
$m$, $k$ and $l$ are parameters depending on the subspace of the H-atom
for which the basis functions have been optimized and
$\alpha$ and $\beta$ are variational parameters.
We remark that the nonlinear optimization of the variational parameters $\alpha$ and $\beta$ 
has to be accomplished for typically of the order of $100$ atomic orbitals and is done by
reproducing many excited states of the hydrogen atom for each field strength separately.
It represents therefore a tedious and time consuming work which has, however, to be done
with great care in order to obtain precise results for the following molecular
structure calculations.  For a more detailed description of the construction of the molecular
electronic wave function we refer the reader to Ref. \cite{detmer1:1997}.

In order to determine the molecular electronic wave function of $H_2$
we use the variational principle which means that we minimize the variational integral
$\frac{\int \Psi^* H \Psi}{\int \Psi^* \Psi}$ by varying the
coefficients $c_i$.
The resulting generalized eigenvalue problem reads as follows:
\begin{equation}
\left(\underline{H} - \epsilon\underline{S}\right)\bbox{c} = \bbox{0} \label{form3}
\end{equation}
where the Hamiltonian matrix $\underline{H}$ is real and symmetric and the overlap matrix
is real, symmetric and positive definite. The vector $\bbox{c}$ contains the expansion coefficients.
The matrix elements of the Hamiltonian matrix and the overlap matrix
are certain combinations of matrix elements with respect to
the optimized nonspherical Gaussian atomic orbitals.
A description of the techniques necessary for the evaluation of these
matrix elements is given in Ref. \cite{detmer1:1997}. We mention
here only that the electron-electron integrals needed a combination of numerical and analytical 
techniques in order to make its rapid evaluation possible. The latter represents the CPU time dominating
factor for the construction of the Hamiltonian matrix.

For the numerical solution of the eigenvalue problem (\ref{form3}) we
used the standard NAG library.
The typical dimension of the Hamiltonian matrix for each subspace varies between
approximately 2000 and 5000 depending on the magnetic field strength.
Depending on the dimension of the Hamiltonian matrix, it takes between 70 and 250 minutes for
simultaneously calculating one point of a PEC of each subspace on a IBM RS6000 computer.
The overall accuracy of our results with respect to the total energy is estimated to be typically
of the order of magnitude of $10^{-4}$ and for some cases of the order of magnitude of $10^{-3}$.
It should be noted that this estimate is rather conservative;
in some ranges of the magnetic field strength and internuclear distance,
e.g., close to the separated atom limit, the accuracy is $10^{-5}$ or even better.
The positions, i.e. internuclear distances, of the maxima and
the minima in the PECs were determined with an accuracy
of $10^{-2} a.u.$ Herefore about 350 points were calculated on an average for each PEC.
It was not necessary to further improve this accuracy
since a change in the internuclear distance about $1\times10^{-2}\;a.u.$ results
in a change in the energy which is typically of the order of magnitude of $10^{-4}$ or smaller.
The total CPU time needed to complete the present work amounts to several years on the 
above powerful computer.

\section{Results and Discussion}

To understand the influence of the external magnetic field on the electronic
structure of the hydrogen molecule we first have to remind ourselves of
the properties in the absence of the field. Accurate data for hydrogen are
of great importance both in astrophysics as well as laboratory physics.
It is a paradigm for many molecular phenomena like charge transfer,
excitation, ionization or scattering processes. Indeed our CI calculations
on the basis of an anisotropic Gaussian basis set provided also 
significant progress with respect to the knowledge of the field-free 
excitations of the molecule: several highly excited states have been
calculated for the first time and some of the PECs for the
lower lying states have been improved. The corresponding results have been
presented to some detail in ref.\cite{detmer2:1998} and contain elaborate information
on the first eight excited singlet and triplet states for both gerade and
ungerade parity. In the following we will first summarize the main properties of the
excited singlet states in the absence of the field and then investigate the electronic
structure in the presence of the magnetic field with increasing field strength.
We hereby first deal with the gerade and subsequently with the ungerade states. 

\subsection{Excited gerade singlet states}

\subsubsection{Field-free states}

The investigation of the electronic states and PECs has been
done for all internuclear distances considered ($0.8\le R \le 1000 a.u.$) with the
same atomic orbital basis set. The latter has been optimized to yield precise
energies (accuracy $10^{-6}-10^{-9}$) of the hydrogen atom for the six lowest
states for both parities for vanishing atomic magnetic quantum number. Additionally,
in order to describe correlation effects, we have included basis functions
with atomic magnetic quantum numbers $1\le m_a \le 5$. The approximate number of
two-particle configurations resulting from the above basis set is $3800$. 
The accuracy of the electronic energies for the higher excited states $n^{1}\Sigma^{+}_{g}, n=7-9$
($n$ indicates the degree of excitation) is, due to the above choice
of the optimized basis set, lower than that for less excited states.
Figure 1 shows the PECs for the states $n^{1}\Sigma^{+}_{g}, n=2-9$
where the dotted lines represent the curves for the higher excited states $n=7-9$. 
There is a large energetical gap ($0.3-0.4 a.u.$) between the ground and the
excited states of $^{1}\Sigma^{+}_{g}$ symmetry. The first five excited states
are well-known from the literature \cite{wolniewicz:1993}. Our calculations
\cite{detmer2:1998} show in most cases an agreement within $10^{-6}-5\cdot10^{-5}$ compared to 
the literature and in several cases also a variationally lower energy. As already 
mentioned the results on the higher excited states ($n=7-9$) have for the first time
been reported very recently in ref.\cite{detmer2:1998}.

As can be seen from figure 1 all the PECs of the states $n^{1}\Sigma^{+}_{g}, n=2-9$
possess a deep potential well around a minimum located approximately at $R=2a.u.$.
A particular feature occuring for most of the considered $^1\Sigma_{g(u)}$ states is the existence
of a second outer minimum and therefore the corresponding PECs exhibit
a double well. Vibrational states in these outer wells \cite{wolniewicz:1993} attracted recently significant experimental interest
\cite{reinhold:1997} since they allow the experimental observation of long-lived and highly excited valence
states of the hydrogen molecule. The two minima of the $3^1\Sigma^+_g$ state arise due to the 
fact that two different configurations of the same symmetry, namely the $1\sigma_g 3d\sigma_g$
and the $1\sigma_g 2s\sigma_g$ configurations, are energetically minimized at two significantly
different internuclear distances. The deep outer wells of the $n^1\Sigma^+_g, n=2,4,7$ states
arise due to a series of avoided crossings between the Heitler-London configurations $H(1s)+H(nl)$
and the ionic configurations $H^+-H^-(1s^2)$. Particularly the $7^1\Sigma^+_g$ state possesses
a very broad and deep ($0.015473a.u.$!) outer potential well which is separated by a broad barrier from the inner
well located at $R\approx 2a.u.$. The outer minimum is located at $R\approx 33.7a.u.$.
A series of avoided crossings at very large internuclear distances $R \approx 300a.u.$ 
leads to the energetically equal dissociation limits $H(1s)+H(4l)$ of the $n^1\Sigma^+_g, n=7-9$
states. The dissociation channel of the $10^1\Sigma^+_g$ state is the ionic configuration $H^++H^-(1s^2)$.
Tables 1 to 3 contain (among the data in the presence of a magnetic field)
the total and dissociation energies at the equilibrium internuclear
distances, the equilibrium internuclear distances and the total energies in the dissociation limit
for the first to third excited $^1\Sigma^{+}_{g},n=2-4$ states in the absence of the magnetic field.

\subsubsection{Evolution in the presence of a magnetic field}

The subspace of $^1\Sigma_g$ symmetry contains the electronic ground state of the hydrogen
molecule in field-free space and in the presence of a magnetic field in the regime
$0<B<0.1a.u.$ For a detailed discussion of the appearance of this state and the global
ground state with increasing field strength in general we refer the reader to refs.\cite{lai:1992,
ortiz:1995,kravchenko1:1997,kravchenko1:1998,detmer1:1997,detmer1:1998} (see also
introduction of the present work). In the following we investigate the evolution of the
excited $n^1\Sigma_g,n=2-8$ states with increasing magnetic field strength for the 
regime $0<B<100a.u.$. We will first study the changes of the PECs
of individual states with increasing field strength and thereafter we present a global
view of the evolution of the spectrum. In order to compare the PECs
for the same state for different field strengths we subtract from the total energies
the corresponding energies in the dissociation limit (which is different for different
field strengths), i.e. we show the quantity $E(R)=E_t(R)-lim_{R\rightarrow \infty} E_t(R)$.
In general the dissociation limit of a certain state of $\Sigma$ symmetry changes
with increasing field strength which is due to the reordering of the energy levels 
of the atoms (hydrogen, hydrogen negative ion) in the external field.
For the atomic states we will use in the following the notation $n m_a^{\pi_a}$
where $n$ specifies the degree of excitation and $m_a,\pi_a$ the atomic magnetic
quantum number and z-parity, respectively.

Let us begin our investigation of the evolution of individual states with 
increasing field strength with the $2^1\Sigma_g$ state whose PECs
are shown in Figure 2a. The positions of the two minima and the corresponding maximum
decrease with increasing field strength. The depth of the inner potential well
decreases for $B \lesssim 0.5 a.u.$ and increases rapidly for $B \gtrsim 1a.u.$.
The depth of the outer well is monotonically increasing for the complete regime
$0<B<100a.u.$. For $B\lesssim 0.01a.u.$ and $B\gtrsim 50a.u.$ the inner well is therefore
deeper than the outer well and vice versa for $0.05\lesssim B \lesssim 20.0 a.u.$
(see figure 2a). The dissociative behaviour of the PECs changes
significantly with increasing field strength. The origin of these changes is the 
fact that for $B \lesssim 10a.u.$ the dissociation channel is $H_2 \rightarrow
H(1 0^+) + H(1 0^-)$ whereas for $B \gtrsim 20a.u.$ we have the asymptotic behaviour
$H_2 \rightarrow H^+ + H^-(1 0_s^+)$ (the index $_s$ stands for spin singlet).
The appearance of the ionic configuration as the
dissociation channel for the low-lying electronic $2^1\Sigma_g$ state can be explained
as follows. It is well-known that the hydrogen negative ion possesses infinitely many
bound states in the presence of a magnetic field of arbitrary strength assuming an 
infinite nuclear mass \cite{avron:1981,bezchastnov:1999}.
Certain of these bound states show a monotonically increasing binding energy
with increasing field strength. The latter surpass then more and more of the energy levels belonging
to two hydrogen atoms one being in the global ground state and the other one in the corresponding excited state.
For a sufficiently strong magnetic field we therefore expect the configuration $H^+ + H^-(1 0_s^+)$
to become the dissociation channel particularly for the first excited state of $^1\Sigma_g$ symmetry.
Due to the long range forces the onset of the asymptotic ($R\rightarrow \infty$) behaviour
of the corresponding PECs with the ionic channel ($H^+ + H^-$) is qualitatively different from
the PECs with a neutral dissociation limit ($H+H$). This explains the different
asymptotic behaviour of the PECs shown in figure 2a with increasing field
strength. Finalizing the discussion of the $2^1\Sigma_g$ state we remark that its PECs
possesses a second maximum for $0.01\lesssim B \lesssim 5.0a.u.$ which however
occurs at large internuclear distances ($R\approx 20a.u.$) and is only of the order of
$10^{-4}a.u.$ above the dissociation limit. Table 1 contains relevant data of the 
PECs of the $2^1\Sigma_g$ state with increasing field strength. 

Next we turn to the second excited i.e. the $3^1\Sigma_g$ state whose PECs
are shown in figure 2b. The positions of the two minima and the corresponding maximum
already present in field-free space decrease monotonically with increasing field strength. 
Starting from $B=0a.u.$ the depth of the inner well decreases with increasing field strength 
whereas it increases for $B \gtrsim 0.5a.u.$. Besides a very small interval
of field strengths the depth of the outer well increases with increasing field strength.
For $B \le 5.0a.u.$ the outer well is deeper than the inner one whereas for $B \gtrsim 10a.u.$
the deep inner well dominates the shape of the PEC. We remark that the
curvature at the (first) maximum and the outer minimum increases significantly with increasing field strength.
The evolution of these increasingly sharper turns can only be fully understood if one looks at the complete
spectrum (see figure 3 and in particular 3(e)) with increasing field strength: 
they develop due to a number of narrow avoided crossing of the first to third excited states
in strong fields. An interesting property of the PEC of the $3^1\Sigma_g$ state is the existence of an additional 
outer (third) minimum for the interval $0.01 \lesssim B \lesssim 10a.u.$ which is
shown in figure 2c. This minimum arises due to the interaction with the ionic configuration $H^+ + H^-$.
In field-free space the lowest and only bound ionic channel $H^+ + H^-(1 0_s^+)$ is the dissociation
channel of the $10^1\Sigma_g^+$ state. With increasing field strength the hydrogen negative 
ion becomes increasingly stronger bound (see discussion above) and therefore it occurs as
the dissociation channel for the sequence of excited states $9^1\Sigma_g,8^1\Sigma_g,...$
finally becoming the dissocation channel of the $2^1\Sigma_g$ state for $B \gtrsim 20a.u.$. 
The existence of the additional outer minimum becomes now understandable:
due to the energetical lowering of the ionic dissociation channel with increasing field strength
the higher excited states of $^1\Sigma_g$ symmetry evolve outer minima and corresponding wells
for certain regimes of the field strength. For the $3^1\Sigma_g$ state this outer well is 
extremely shallow for $B\lesssim 0.05a.u.$ and therefore almost invisible in figure 2c.
For $B \gtrsim 0.1a.u.$ it becomes however well-pronounced. Between $B=0.5a.u.$ and $B=1.0a.u.$
there occurs a change with respect to the dissociation channel of the $3^1\Sigma_g$. For
$0<B\lesssim 0.5a.u.$ the dissociation channel is $H(10^+) + H(2 0^+)$ and for $1.0\lesssim B
\lesssim 10a.u.$ it is $H^+ + H^-(0_s^+)$. The similar asymptotic behaviour of the
PECs belonging to different field strengths (see figure 2c for $B=1.0, 5.0$ and $10.0$) arises due to the
fact that they possess all the ionic dissociation channel. In the latter regime the position
of the (third) outer minimum increases with increasing field strength 
(for $B=10.0a.u.$ the outer minimum is located at $\approx 85a.u.$). Finally there is a
second change of the dissociation channel of the $3^1\Sigma_g$ state to $H(1 0^+) + H(1 0^-)$
and therefore the outer minimum disappears for $B \gtrsim 20a.u.$. 
Table 2 contains the total and dissociation energies at the equilibrium internuclear
distances, the equilibrium internuclear distances and the total energies in the dissociation limit
for the second excited $3^1\Sigma^{+}_{g}$ state in the regime $0<B<100a.u.$

Next we focus on the third excited $4^1\Sigma_g$ whose PECs with
increasing field strength are shown in figure 2d. In field-free space it possesses two
minima and associated potential wells located at $R=1.97a.u.$ and $R=11.21a.u.$, respectively.
The position of the inner minimum increases with increasing
field strength whereas the corresponding dissociation energy decreases. Finally for $B \approx
0.2a.u.$ the associated well disappears but reappears for $B \gtrsim 0.5 a.u.$.
With further increasing field strength the position of this inner minimum 
decreases and the depth of the corresponding well increases monotonically
for $B \gtrsim 1a.u.$. Independently of this first inner minimum and the outer minimum
there appears for $B\gtrsim 0.2a.u.$ an additional third minimum and corresponding 
well (see table 3 and figure 2d) for small internuclear distances $1-3a.u.$. Although this
new minimum and well are energetically well below the dissociation limit for $B \gtrsim 20.0a.u.$
they are separated from the other inner minimum only by a tiny barrier. These facts will become
better understandable in the context of our discussion of the evolution of the whole
spectrum with increasing field strength (see below). The properties of the PEC of the
$4^1\Sigma_g$ state at large internuclear distances are somewhat analogous to that
of the $3^1\Sigma_g$ state. The outer minimum has its origin in the interaction of the
neutral $H+H$ and ionic $H^++H^-$ configurations. Starting with $B=0$ and increasing the field strength
the depth of the outer well increases. The first change of the dissociation channel from
$H_2\rightarrow H(1 0^+) + H(3 0^+)$ to $H_2\rightarrow H^+ + H^-(1 0^+_s)$
occurs in the regime $0.1<B<0.2a.u.$. In the regime $0.1\lesssim B \lesssim 1.0a.u.$
the position of the outer minimum increases with increasing field strength (for $B=0.5a.u.$ it is
already $R=50a.u.$) and the depth of the outer well decreases.  
Due to the further increasing binding energy of the hydrogen negative ion $1 0^+_s$ state with increasing field strength
we encounter a second change of the dissociation channel at $B \approx 1.0a.u.$ to $H_2 \rightarrow H(1 0^+) + H(2 0^+)$
which causes the disappearance of the outer minimum and well. Table 3 provides the
corresponding data for the $4^1\Sigma_g$ state.

The PECs of the $5^1\Sigma_g$ and $6^1\Sigma_g$ states are shown in figures 2e and 2f, respectively.
For both states the positions of the maxima and minima as well as the corresponding total energies
show an 'irregular' behaviour as a function of the field strength for $B \lesssim 2.0a.u.$.
We therefore focus on the main features of these states. For certain regimes of the field strength
we observe double well structures for the PECs. Analogously to the $n^1\Sigma_g, n=2-4$ states
there exist additional outer minima and wells due to the interaction with the ionic configuration
for certain field strength regimes. For $B > 2.0a.u.$ the position of the first inner minimum
decreases rapidly with increasing field strength whereas the corresponding dissociation energy 
increases. Also we observe the existence of minima whose energies lie above the dissociation
energy, i.e. the corresponding wells contain if at all metastable states.
We remark that some of the above-discussed features, in particular those associated with small energy
scales, might not be visible in the corresponding figures 2 but only in a zoom of the
relevant regimes of internuclear distances of the considered PECs.
We again emphasize that due to the large amount of data we do not present full PECs or
data on the higher excited states $n^1\Sigma_g, n=5-8$ which can be obtained from the
authors upon request.
 
\subsubsection{Discussion of the evolution of the complete spectrum}

In the present subsection we focus on the evolution of the complete spectrum of the
excited $n^1\Sigma_g, n=2-8$ states with increasing field strength. This will give us the complementary
information to the evolution of individual states presented above. Figure 3a-f shows the corresponding
PECs for the field strengths $B=0.01, 0.1, 0.5, 5.0, 100.0a.u.$, respectively.
The PECs of the five energetical lowest excited states $n^1\Sigma_g, n=2-6$ are hereby illustrated with
full lines indicating their higher accuracy whereas the PECs of the electronic states $n^1\Sigma_g, n=7,8$
are less accurate and illustrated with dotted lines. Before we discuss the evolution with increasing
field strength some general remarks are in order. The energy gap between the ground state $1^1\Sigma_g$
and the first excited state $2^1\Sigma_g$ is of the order of $0.4a.u.$ in field-free space and increases
montonically with increasing field strength. At the same time the total energies of all states $n^1\Sigma_g$
are shifted in lowest order proportional to $B$ with increasing field strength which is due to the
raise of the kinetic energy in the presence of a magnetic field. 

In field-free space many of the dissocation channels of the PECs of excited $^1\Sigma_g$ states are degenerate
due to the degeneracies of the field-free hydrogen atom (see figure 1 and 4).
The major difference of the PECs in field-free space compared to those for weak fields 
is the removal of these degeneracies (see, for example, figure 3a for $B=0.01a.u.$). With increasing
field strength figures 3a-d ($B=0.01,0.1,0.5~and~5.0a.u.$) demonstrate the systematic lowering of the
diabatic energy curve belonging to the ionic configuration $H^++H^-(1s^2)$. This diabatic curve passes through
the spectrum with increasing field strength thereby causing an intriguing evolution of avoided crossings and corresponding
potential wells for the individual states. At $B\approx 0.1a.u.$ the fourth excited $5^1\Sigma_g$ state 
acquires the ionic dissociation channel. The $7^1\Sigma_g$ state thereby looses its outer potential well
which was very well-pronounced in the absence of the external field. In the same course the $3^1\Sigma_g$ state
shows a number of avoided crossings with the $2^1\Sigma_g$ state:
it develops an additional outer minimum and well which is rather deep at $B\approx 0.5a.u.$ accompanied by the
flattening of the first inner well and the deepening of the second inner well.
Furthermore we observe for $B\approx 0.5a.u.$ the appearance of a large
number of avoided crossing among the higher excited states $n^1\Sigma_g, n=5-8$ at $R\approx 5a.u.$.
At $B\approx 0.2a.u.$ the third excited $4^1\Sigma_g$ state acquires the ionic dissociation channel.
Subsequently, i.e. with further increasing field strength, the second excited $3^1\Sigma_g$ state 
(see figure 3d) and finally the first excited $2^1\Sigma_g$ state acquire ionic character for sufficiently large
internuclear distances. In the high field situation (see figure 3e for $B=100a.u.$) only the energetically lowest
excited state possess a well-pronounced double well structure and the overall picture is dominated
by the fact that the PECs of the considered states possess a  very similar shape and are energetically very close to each other 
in particular around the inner minimum at small internuclear
distances. Figure 3f shows for $B=100a.u.$ a zoom of the series of avoided crossings occuring for
the higher excited states $n^1\Sigma_g,n=5-10$ in the regime $2<R<12a.u.$.

\subsection{Excited ungerade singlet states}

\subsubsection{Field-free states}

The four energetically lowest states of $^1\Sigma^+_u$ symmetry at $B=0$ have been investigated in
detail and with high accuracy in the literature \cite{dressler:1995}. Our results \cite{detmer2:1998} show
a relative accuracy of $10^{-4}$ for the energies of the $1^1\Sigma^+_u$ state and of
$10^{-5}$ for the first two excited states i.e. the $n^1\Sigma^+_u, n=2,3$ states.
The energies of the $4^1\Sigma^+_u$ state are significantly lower than the data presented 
in \cite{dressler:1995}. The PECs for the $n^1\Sigma^+_u, n=4-9$ presented
in ref.\cite{detmer2:1998} for the first time are estimated to possess an accuracy of
$10^{-5}$ for the $n^1\Sigma^+_u, n=4-6$ states and $10^{-4}$ for the $n^1\Sigma^+_u, n=7-9$ states.
Figure 4 shows the PECs of the ground as well as eight excited
states of $^1\Sigma^+_u$ symmetry in the range $1<R<1000a.u.$ on a logarithmic scale.

The PEC of the ground state $1^1\Sigma^+_u$ of ungerade symmetry
possesses a minimum at $R=2.43a.u.$ and a corresponding deep well. A closer look at the
wave function reveals its ionic character for $3<R<7a.u.$. With further increasing internuclear
distance the ionic character of the wave function decreases and the corresponding
dissociation channel is $H_2 \rightarrow H(1s)+H(2p)$. The PEC of the
first excited $2^1\Sigma^+_u$ state is similar to that of the ground state $1^1\Sigma^+_u$:
its equilibrium internuclear distance is $R_{eq}=2.09$ the dissociation channel is identical to that
of the $1^1\Sigma^+_u$ state. The depth of its single well is however only one third of
the depth of the well of the $1^1\Sigma^+_u$ state. For the higher excited states we observe
a similar behaviour as in the case of the excited electronic states of $^1\Sigma_g^+$ symmetry.
The PECs of the $n^1\Sigma^+_u, n=3-9$ states possess a deep well around
a minimum located approximately at $R \approx 2a.u.$. Furthermore the $n^1\Sigma^+_u, n=3,6$
states exhibit additional deep outer potential wells at large internuclear distances which 
arise due to the avoided crossings of the Heitler-London configurations with the
corresponding ionic configuration. The outer minimum of the $6^1\Sigma^+_u$ state
is located at $33.7a.u.$ and the corresponding well possesses a remarkable depth of $0.015134a.u.$:
it is expected to contain a large number of long-lived vibrational states.
Tables 4 to 6 contain (among the data in the presence of the magnetic field)
the total and dissociation energies at the equilibrium internuclear
distances, the equilibrium internuclear distances and the total energies in the dissociation limit
for the first to third excited $^1\Sigma^{+}_{u}$ states in the absence of the magnetic field.

\subsubsection{Evolution in the presence of a magnetic field}

First of all we remark that the dissociation channels of the $(n+1)^1\Sigma_g$ states coincide with
those of the $n^1\Sigma_u$ states for $n=1-7$ in the complete regime $0\le B \le 100a.u.$.
The qualitative behaviour of the PECs of the $n^1\Sigma_u$ states at large internuclear distances is
therefore similar to that of the $(n+1)^1\Sigma_g$ states discussed in the previous section.
In particular many of the explanations and remarks provided there hold also for the present case 
of the $n^1\Sigma_u$ states.

Before discussing the behaviour of the PECs of the individual excited $n^1\Sigma_u$ states with increasing field 
strength some remarks concerning the lowest, i.e. ground state of $^1\Sigma_u$ symmetry are in order
(for its PEC with increasing field strength see figure 6).
Its dissociation energy increases monotonically with increasing field strength whereas its 
equilibrium internuclear distance increases slightly for weak fields and decreases significantly for
increasingly stronger fields. As indicated above the asymptotic $R\rightarrow \infty$ behaviour of the PECs of the
$1^1\Sigma_u$ and $2^1\Sigma_g$ states is very similar. For $B=100a.u.$ the PEC of the $1^1\Sigma_u$ state possesses
a peculiar shape which is largely determined by the ionic dissociation channel $H^+ + H^-(10_s^+)$
(see figure 6f). For more details on this state we refer the reader to ref.\cite{detmer1:1997,detmer1:1998}.

The first excited $2^1\Sigma_u$ state possesses in field-free space an equilibrium
internuclear distance $R_{eq}=2.09a.u.$. Figure 5a shows the corresponding PEC with increasing
field strength for $0 \le R \le 5a.u.$ whereas figure 5b illustrates particularly the behaviour
at large internuclear distances. In the regime $0 \le B \lesssim 0.2a.u.$ the dissociation 
energy decreases slightly and the bond length increases. With further increasing field strength
the dissociation energy increases drastically and the bond length decreases. 
For $0.1 \lesssim B \lesssim 50a.u.$ there exists a maximum and a corresponding additional outer minimum
at large internuclear distances (see figure 5b) whose origin is again the emergence
of the ionic configuration for the wave function of the $2^1\Sigma_u$ state.
Figure 5b also demonstrates the similarity of the asymptotic $R \rightarrow \infty$
behaviour of the PECs of the $2^1\Sigma_u$ state
in the regime $1.0 \le B \le 10.0a.u.$. The corresponding data for the PECs of the first excited $2^1\Sigma_u$ state
are given in table 4.

Turning to the second excited $3^1\Sigma_u$ state we observe that the depth of the
potential well located for $B=0a.u.$ at $R_{eq}=2.03a.u.$ decreases for weak fields whereas
it increases significantly for strong fields $B \gtrsim 1.0a.u.$ (see figure 5c). 
The existence of an additional outer minimum for this state can be seen in figure 5d.
In many respects a similar behaviour to that of the $2^1\Sigma_u$ state is observed
although, of course, the regimes of field strength for which the individual phenomena
take place are different. Table 5 contains the corresponding data of the PECs of the $3^1\Sigma_u$ state.
Finally figures 5e and 5f show the PECs of the $n^1\Sigma_u, n=4,5$ states
with increasing field strength, respectively. They exhibit a number of maxima and minima most
of which can however hardly be seen in figures 5e,f or occur at large internuclear distances.
The origin of their existence are again the different (ionic and neutral) dissociation channels.
These maxima and minima are present only for certain individually different regimes of the field strength.
Some of them are located above and some of them below the dissociative threshold.
As can be seen the bond length (belonging to the inner minimum) decreases monotonically
and the dissociation energy increases significantly above some critical value $B_c$.
The inner minimum and associated well possesses a remarkably large dissociation energy for strong fields.
Table 6 provides data on the PECs of the $4^1\Sigma_u$ state.

To finalize our discussion on the $^1\Sigma_u$ subspace we show in figure 6 the evolution of the
spectrum with increasing field strength. Figures 6a-f show the PECs for the $n^1\Sigma_u,n=1-8$ states
for the field strengths $B=0.05, 0.1, 0.5, 1.0, 10.0, 100.0a.u.$, respectively.
Analogously to the case of the $^1\Sigma_g$ subspace we observe for weak fields the removal of the 
degeneracies due to the field-free hydrogen atom in the dissociation limit.
With increasing field strength we see the lowering of the diabatic energy line belonging to the
ionic configuration which causes the appearance and disappearance of outer maxima, minima and 
corresponding outer potential wells until finally ($B=100a.u.$) the $1^1\Sigma_u$ state possess
the ionic dissociation channel $H_2 \rightarrow H^+ + H^-(10_s^+)$ which is the origin of the 
peculiar shape of its PEC. A number of further observations made for the manifold of the $n^1\Sigma_g,n=1-8$ states
above can also be seen for the $n^1\Sigma_u,n=1-8$ states in figure 6 like, for example, 
the similar shape of the potential wells of the excited states in the high field limit.

\section{Conclusions}

The hydrogen molecule is the most fundamental molecular system and of immediate importance
in a variety of different physical circumstances. In spite of the fact that it has been investigated
over the past decades in great detail and that our knowledge on this system has grown enormously
there are plenty of questions and problems to be addressed even for the molecule in field-free space.
As an example we mention certain highly excited Rydberg states ($7^1\Sigma_g^+,6^1\Sigma_u^+$)
which, due to the ionic character of the
binding for certain regimes of the internuclear distance, possesses a deep outer well at large
distances which contains a considerable number of vibrational states. On the other hand the detailed
knowledge of hydrogen (even of highly excited states) is of utmost importance for our understanding
and interpretation of the astrophysically observed interstellar radiation.

Much less is known about the behaviour of the hydrogen molecule in strong magnetic fields.
With increasing field strength the ground state of the molecule undergoes two transitions which
are due to a change of the spin and orbital character, respectively. Very recently
the global ground state configurations have been identified for the parallel configuration (there are
good reasons which lead to the conjecture that the derived results hold for arbitrary angle of the
internuclear and magnetic field axis) both on the Hartree-Fock level \cite{kravchenko1:1997,kravchenko1:1998}
and via a fully correlated approach \cite{detmer1:1997,detmer1:1998}. For low fields the ground state
is of spin singlet $^1\Sigma_g$ symmetry, for intermediate fields the spin triplet $^3\Sigma_u$ state
represent the ground state whereas in the high field regime the $^3\Pi_u$ state is the energetically
lowest state. 

The present work goes for the first time beyond the ground state properties and investigates excited
states of the hydrogen molecule in the broad regime $0<B<100a.u.$. We hereby focus on singlet states
of both gerade as well as ungerade symmetry: up to seven excited states have been studied
for the parallel configuration with a
high accuracy of the obtained PECs. A variety of different phenomena have been observed out of
which we mention here only the most important ones. Double
well structures observed in particular for the field-free $n^1\Sigma_g^+$ states are severly modified
in the presence of the field thereby showing a 'coming and going' of new maxima and minima as well
as corresponding wells. The overall tendency in the strong field limit is the development of deep
inner wells containing a large number of vibrational states. In the course of the increasing field strength
a fundamental phenomenon occurs which has a strong impact on the overall shape of the PECs.
Due to the fact that the hydrogen negative ion becomes increasingly bound with increasing
field strength we encounter changes in the dissociation channels of individual states from neutral
$H_2 \rightarrow H + H^*$ to ionic $H_2 \rightarrow H^+ + H^-$ character. For a certain regime of
field strength $B_{c1}<B<B_{c2}$ a certain excited state possesses therefore the ionic dissociation channel thereby
modifying the asymptotic behaviour of its PEC to an attractive Coulombic tail. For weaker fields $B<B_{c1}$ higher excited
states possess this ionic dissociation channel whereas for stronger fields $B>B_{c2}$ it belongs to increasingly
lower excitations. These facts influence the overall appearance of the spectrum thereby
creating features like outer potential wells and/or largely changing avoided crossings. 

The data on the PECs of the excited singlet states obtained here should serve as part of 
the material to be accumulated for the investigation of quasimolecular absorption features
in magnetic white dwarfs. The investigation of excited triplet states of $\Sigma$ symmetry
or of $\Pi$ states, which are of equal importance, are left to future investigations.

\section{Acknowledgment}
Fruitful discussions with W.Becken are gratefully acknowledged. The Deutsche Forschungsgemeinschaft
is gratefully acknowledged for financial support.

\vspace*{2.0cm}

\begin{center}
{\large{\bf FIGURE CAPTIONS}}
\end{center}
\vspace*{1.0cm}

\noindent
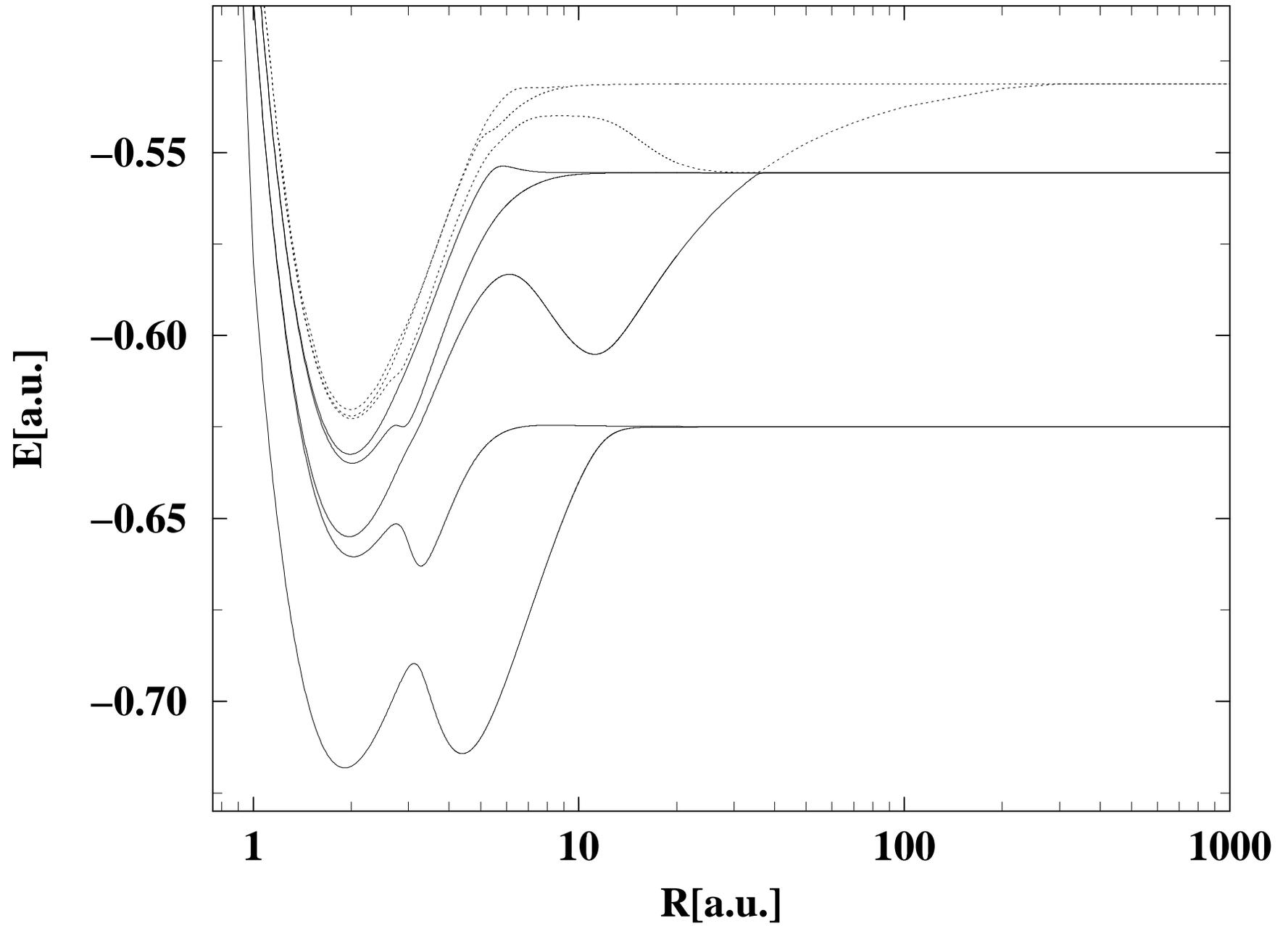
{\bf{Figure 1:}} The potential energy curves of the excited $n^1\Sigma_g^+,n=2-9$ electronic states
of the hydrogen molecule in the absence of a magnetic field.
\vspace*{0.5cm}

\noindent
{\bf{Figure 2:}} The evolution of the potential energy curves for some excited $^1\Sigma_g$ electronic states
of the hydrogen molecule in the presence of a magnetic field $0<B<100a.u.$. In detail are shown the
evolution of the PECs for the (a) $2^1\Sigma_g$, (b) $3^1\Sigma_g$, (c) zoom of $3^1\Sigma_g$,  (d) $4^1\Sigma_g$,
(e) $5^1\Sigma_g$ and (f) $6^1\Sigma_g$ states, respectively.
Shown is the quantity $E(R)=E_t(R)-lim_{R\rightarrow \infty} E_t(R)$ where $E_t(R)$ is the total energy.
\vspace*{0.5cm}

\noindent
{\bf{Figure 3:}} The spectrum of potential energy curves for the excited $n^1\Sigma_g,n=2-8$ electronic states
of the hydrogen molecule in the presence of a magnetic field $0<B<100a.u.$ with increasing field
strength. In detail are shown the PECs for (a) $B=0.01$  (b) $B=0.1$ (c) $B=0.5$ (d) $B=5.0$ 
(e) $B=100.0$  and (f) zoom of $B=100.0a.u.$, respectively.
\vspace*{0.5cm}

\noindent
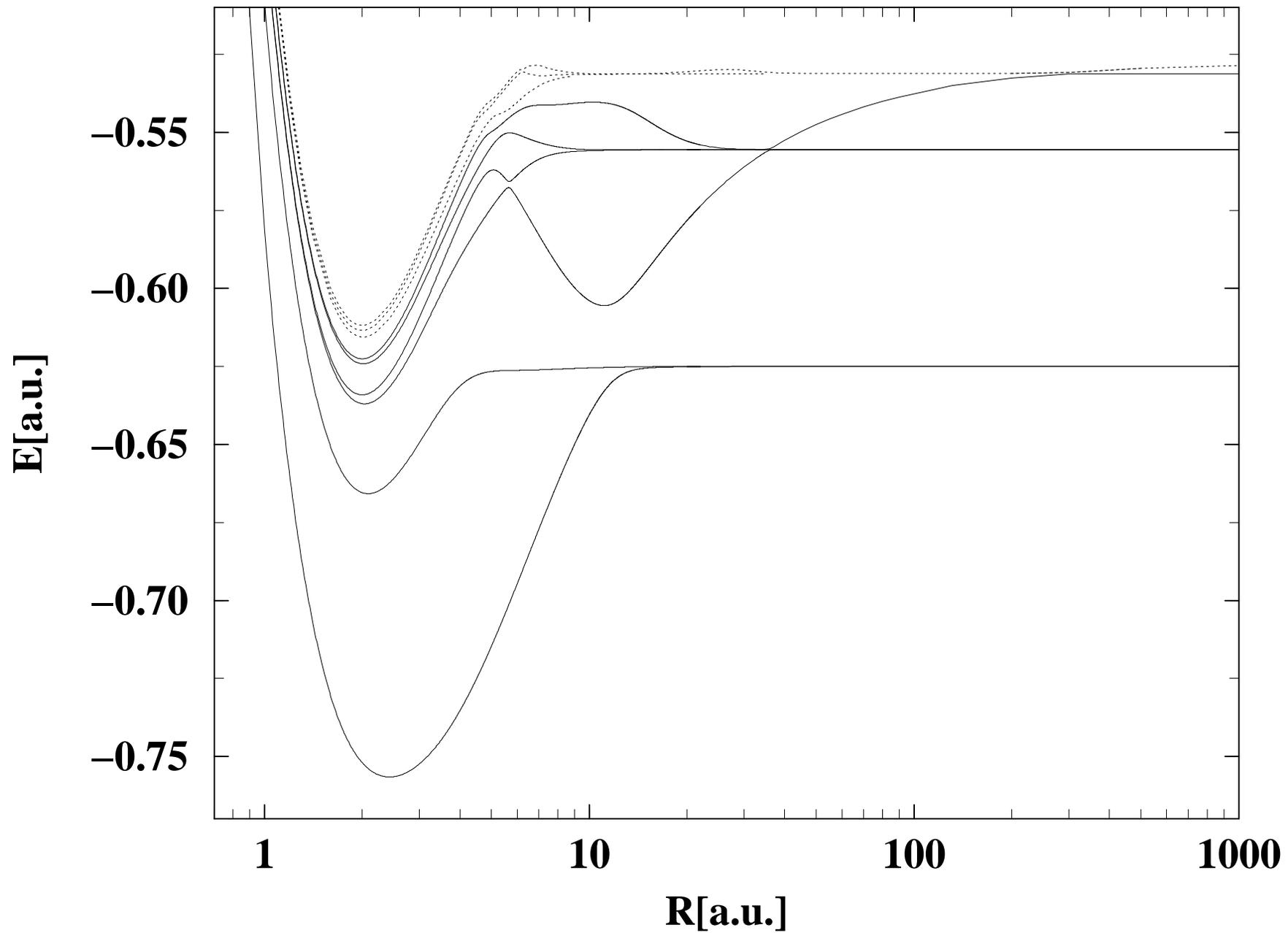
{\bf{Figure 4:}} The potential energy curves of the excited $n^1\Sigma_u^+,n=1-9$ electronic states
of the hydrogen molecule in the absence of a magnetic field.
\vspace*{0.5cm}

\noindent
{\bf{Figure 5:}} The evolution of the potential energy curves for some excited $^1\Sigma_u$ electronic states
of the hydrogen molecule in the presence of a magnetic field $0<B<100a.u.$. In detail are shown the
evolution of the PECs for the (a) $2^1\Sigma_g$, (b) zoom of $2^1\Sigma_g$, (c) $3^1\Sigma_g$,  (d) zoom of $3^1\Sigma_g$,
(e) $4^1\Sigma_g$ and (f) $5^1\Sigma_g$ states, respectively.
Shown is the quantity $E(R)=E_t(R)-lim_{R\rightarrow \infty} E_t(R)$ where $E_t(R)$ is the total energy.
\vspace*{0.5cm}

\noindent
{\bf{Figure 6:}} The spectrum of potential energy curves for the excited $n^1\Sigma_u,n=1-8$ electronic states
of the hydrogen molecule in the presence of a magnetic field $0<B<100a.u.$ with increasing field
strength. In detail are shown the PECs for (a) $B=0.05$  (b) $B=0.1$ (c) $B=0.5$ (d) $B=1.0$ 
(e) $B=10.0$  and (f) $B=100.0a.u.$, respectively.
\vspace*{0.5cm}

\newpage

\begin{center}
\Large{\bf{Tables}}
\end{center}

\begin{table}
\caption{Data for the first excited $^1\Sigma_g$ state: 
Total energies $E_{t1},E_{t2}$ and dissociation energies $E_{d1},E_{d2}$ 
at the corresponding equilibrium internuclear distance, 
the equilibrium internuclear distances $R_{eq1},R_{eq2}$ 
and the total energy in the dissociation limit 
$\lim\limits_{R\to\infty} E_{t}$ as a function of the field strength $0 \leq B \leq 100$ 
(all quantities are given in atomic units).}
\begin{tabular}{dddddddd}
\hline\hline
\multicolumn{1}{c}{\rule[-5mm]{0mm}{11mm}{\raisebox{-0.5ex}[0.5ex]{B}}} & 
\multicolumn{1}{c}{\raisebox{-0.5ex}[0.5ex]{\hspace*{0.5cm} $R_{eq1}$}} &
\multicolumn{1}{c}{\raisebox{-0.5ex}[0.5ex]{\hspace*{0.5cm} $E_{d1}$}} &
\multicolumn{1}{c}{\raisebox{-0.5ex}[0.5ex]{\hspace*{0.5cm} $E_{t1}$}} &
\multicolumn{1}{c}{\raisebox{-0.5ex}[0.5ex]{\hspace*{0.3cm} $R_{eq2}$}} &
\multicolumn{1}{c}{\raisebox{-0.5ex}[0.5ex]{\hspace*{0.5cm} $E_{d2}$}} &
\multicolumn{1}{c}{\raisebox{-0.5ex}[0.5ex]{\hspace*{0.5cm} $E_{t2}$}} &
\multicolumn{1}{c}{\hspace*{0.5cm} $\lim\limits_{R\to\infty} E_{tot}$} \\ \hline
0.0   &  1.91 & 0.093122   & -0.718121 &  4.39 & 0.089241 & -0.714240 & -0.624999 \\
0.001 &  1.91 & 0.093120   & -0.718117 &  4.39 & 0.089242 & -0.714239 & -0.624997 \\
0.005 &  1.91 & 0.093061   & -0.718017 &  4.39 & 0.089252 & -0.714208 & -0.624956 \\
0.01  &  1.91 & 0.092878   & -0.717703 &  4.39 & 0.089289 & -0.714114 & -0.624825 \\
0.05  &  1.90 & 0.087831   & -0.708672 &  4.39 & 0.090459 & -0.711300 & -0.620841 \\
0.1   &  1.88 & 0.077017   & -0.686953 &  4.38 & 0.093325 & -0.703261 & -0.609936 \\
0.2   &  1.88 & 0.057630   & -0.633195 &  4.33 & 0.100709 & -0.676274 & -0.575565 \\
0.5   &  1.89 & 0.040502   & -0.462472 &  4.08 & 0.122427 & -0.544397 & -0.421970 \\
1.0   &  1.78 & 0.044819   & -0.135993 &  3.76 & 0.150618 & -0.241792 & -0.091174 \\
2.0   &  1.54 & 0.067740   &  0.612336 &  3.37 & 0.191218 &  0.488858 &  0.680076 \\
5.0   &  1.18 & 0.140991   &  3.130994 &  2.88 & 0.269224 &  3.002761 &  3.271985 \\
10.0  &  0.95 & 0.245637   &  7.623918 &  2.55 & 0.351756 &  7.517799 &  7.869555 \\
20.0  &  0.76 & 0.359056   & 16.960943 &  2.27 & 0.408351 & 16.911648 & 17.319999 \\
50.0  &  0.56 & 0.575262   & 45.784366 &  1.96 & 0.471572 & 45.888056 & 46.359628 \\
100.0 &  0.44 & 0.821768   & 94.614199 &  1.76 & 0.521822 & 94.914145 & 95.435967 \\
\hline\hline
\end{tabular}
\end{table}

\begin{table}
\squeezetable
\caption{Data for the second excited $^1\Sigma_g$ state: 
Total energies $E_{t1}\,-\,E_{t3}$ and dissociation energies $E_{d1}\,-\,E_{d3}$ 
at the corresponding equilibrium internuclear distance, 
the equilibrium internuclear distances $R_{eq1}\,-\,R_{eq3}$ 
and the total energy in the dissociation limit 
$\lim\limits_{R\to\infty} E_{t}$ as a function of the field strength $0 \leq B \leq 100$ 
(all quantities are given in atomic units).}
{\small
\begin{tabular}{ddddddddddd}
\hline\hline
\multicolumn{1}{c}{\rule[-5mm]{0mm}{11mm}{\raisebox{-0.5ex}[0.5ex]{B}}} & 
\multicolumn{1}{c}{\raisebox{-0.5ex}[0.5ex]{\hspace*{0.2cm} $R_{eq1}$}} &
\multicolumn{1}{c}{\raisebox{-0.5ex}[0.5ex]{\hspace*{0.5cm} $E_{d1}$}} &
\multicolumn{1}{c}{\raisebox{-0.5ex}[0.5ex]{\hspace*{0.5cm} $E_{t1}$}} &
\multicolumn{1}{c}{\raisebox{-0.5ex}[0.5ex]{\hspace*{0.2cm} $R_{eq2}$}} &
\multicolumn{1}{c}{\raisebox{-0.5ex}[0.5ex]{\hspace*{0.5cm} $E_{d2}$}} &
\multicolumn{1}{c}{\raisebox{-0.5ex}[0.5ex]{\hspace*{0.5cm} $E_{t2}$}} &
\multicolumn{1}{c}{\raisebox{-0.5ex}[0.5ex]{\hspace*{0.0cm} $R_{eq3}$}} &
\multicolumn{1}{c}{\raisebox{-0.5ex}[0.5ex]{\hspace*{0.5cm} $E_{d3}$}} &
\multicolumn{1}{c}{\raisebox{-0.5ex}[0.5ex]{\hspace*{0.5cm} $E_{t3}$}} &
\multicolumn{1}{c}{\hspace*{0.5cm} $\lim\limits_{R\to\infty} E_{tot}$} \\ \hline
0.0   &  2.03 & 0.035466   & -0.660465 &  3.27 & 0.038014 & -0.663013 &       &          &           & -0.624999 \\
0.001 &  2.03 & 0.035462   & -0.660458 &  3.27 & 0.038021 & -0.663017 &       &          &           & -0.624996 \\
0.005 &  2.03 & 0.035399   & -0.660305 &  3.27 & 0.038045 & -0.662951 &       &          &           & -0.624906 \\
0.01  &  2.03 & 0.035225   & -0.659851 &  3.27 & 0.038101 & -0.662727 & 18.97 & 0.000007 & -0.624633 & -0.624626 \\
0.05  &  2.04 & 0.033199   & -0.649592 &  3.20 & 0.040478 & -0.656871 & 10.56 & 0.001510 & -0.617903 & -0.616393 \\
0.1   &  2.03 & 0.030981   & -0.626596 &  3.06 & 0.048089 & -0.643704 & 10.02 & 0.006992 & -0.602607 & -0.595615 \\
0.2   &  1.94 & 0.023894   & -0.563261 &  2.86 & 0.064882 & -0.604249 & 10.43 & 0.026352 & -0.565719 & -0.539367 \\
0.5   &  1.90 & 0.030305   & -0.378302 &  2.73 & 0.079773 & -0.427788 & 10.69 & 0.066892 & -0.414907 & -0.348015 \\
1.0   &  1.78 & 0.039842   & -0.041327 &  2.50 & 0.091331 & -0.092816 & 10.96 & 0.085978 & -0.087463 & -0.001485 \\
2.0   &  1.54 & 0.042657   &  0.717581 &  2.22 & 0.089173 &  0.671065 & 12.14 & 0.079131 &  0.681107 &  0.760238 \\
5.0   &  1.19 & 0.075290   &  3.248431 &  1.86 & 0.091565 &  3.232156 & 19.38 & 0.051644 &  3.272077 &  3.323721 \\
10.0  &  0.96 & 0.131036   &  7.749942 &  1.63 & 0.093390 &  7.787588 & 85.58 & 0.011423 &  7.869555 &  7.880978 \\
20.0  &  0.76 & 0.275591   & 17.095639 &  1.44 & 0.142076 & 17.229154 &       &          &           & 17.371230 \\
50.0  &  0.56 & 0.605249   & 45.931209 &  1.22 & 0.250618 & 46.285840 &       &          &           & 46.536458 \\
100.0 &  0.45 & 0.976059   & 94.770524 &  1.09 & 0.357751 & 95.388832 &       &          &           & 95.746583 \\
\hline\hline
\end{tabular}
}
\end{table}

\begin{table}
\squeezetable
\caption{Data for the third excited $^1\Sigma_g$ state: 
Total energies $E_{t1}\,-\,E_{t3}$ and dissociation energies $E_{d1}\,-\,E_{d3}$ 
at the corresponding equilibrium internuclear distance, 
the equilibrium internuclear distances $R_{eq1}\,-\,R_{eq3}$ 
and the total energy in the dissociation limit 
$\lim\limits_{R\to\infty} E_{t}$ as a function of the field strength $0 \leq B \leq 100$ 
(all quantities are given in atomic units).}
{\small
\begin{tabular}{ddddddddddd}
\hline\hline
\multicolumn{1}{c}{\rule[-5mm]{0mm}{11mm}{\raisebox{-0.5ex}[0.5ex]{B}}} & 
\multicolumn{1}{c}{\raisebox{-0.5ex}[0.5ex]{\hspace*{0.0cm} $R_{eq1}$}} &
\multicolumn{1}{c}{\raisebox{-0.5ex}[0.5ex]{\hspace*{0.5cm} $E_{d1}$}} &
\multicolumn{1}{c}{\raisebox{-0.5ex}[0.5ex]{\hspace*{0.5cm} $E_{t1}$}} &
\multicolumn{1}{c}{\raisebox{-0.5ex}[0.5ex]{\hspace*{0.0cm} $R_{eq2}$}} &
\multicolumn{1}{c}{\raisebox{-0.5ex}[0.5ex]{\hspace*{0.5cm} $E_{d2}$}} &
\multicolumn{1}{c}{\raisebox{-0.5ex}[0.5ex]{\hspace*{0.5cm} $E_{t2}$}} &
\multicolumn{1}{c}{\raisebox{-0.5ex}[0.5ex]{\hspace*{0.0cm} $R_{eq3}$}} &
\multicolumn{1}{c}{\raisebox{-0.5ex}[0.5ex]{\hspace*{0.5cm} $E_{d3}$}} &
\multicolumn{1}{c}{\raisebox{-0.5ex}[0.5ex]{\hspace*{0.5cm} $E_{t3}$}} &
\multicolumn{1}{c}{\hspace*{0.3cm} $\lim\limits_{R\to\infty} E_{tot}$} \\ \hline
0.0   &  1.97 & 0.099408    & -0.654963 & \imina&  \iminb  & \iminb    & 11.21 & 0.049597  & -0.605152 & -0.555555 \\
0.001 &  1.97 & 0.099397    & -0.654944 & \imina&  \iminb  & \iminb    & 11.21 & 0.049604  & -0.605151 & -0.555547 \\
0.005 &  1.97 & 0.099051    & -0.654469 & \imina&  \iminb  & \iminb    & 11.21 & 0.049686  & -0.605104 & -0.555418 \\
0.01  &  1.97 & 0.098011    & -0.653029 & \imina&  \iminb  & \iminb    & 11.21 & 0.049930  & -0.604948 & -0.555018 \\
0.05  &  1.99 & 0.081038    & -0.625896 &  2.93 & 0.068063 & -0.612921 & 11.39 & 0.055502  & -0.600360 & -0.544858 \\
0.1   &  2.05 & 0.075239    & -0.597699 & \imina&  \iminb  & \iminb    & 12.04 & 0.06419  & -0.586650 & -0.522460 \\
0.2   & \imina&  \iminb     & \iminb    &  2.46 & 0.067543 & -0.546109 & 15.62 & 0.059554  & -0.538120 & -0.478566 \\
0.5   &  1.91 & 0.022150    & -0.350029 &  2.35 & 0.035929 & -0.363808 & 50.06 &  0.020138   & -0.348017 & -0.327879 \\
1.0   &  1.77 & 0.019181    & -0.010815 &  2.17 & 0.030750 & -0.022384 & \imina&  \iminb     &  \iminb   &  0.008366 \\
2.0   &  1.53 & 0.053756    &  0.750088 &  1.95 & 0.058173 &  0.745671 & \imina&  \iminb     &  \iminb   &  0.803844 \\
5.0   &  1.19 & 0.142822    &  3.283035 &  1.67 & 0.116673 &  3.309184 & \imina&  \iminb     &  \iminb   &  3.425857 \\
10.0  &  0.96 & 0.257211    &  7.786043 &  1.50 & 0.176832 &  7.866422 & \imina&  \iminb     &  \iminb   &  8.043254 \\
20.0  &  0.76 & 0.427463    & 17.133303 &  1.34 & 0.249318 & 17.311448 & \imina&  \iminb     &  \iminb   & 17.560766 \\
50.0  &  0.56 & 0.768452    & 45.971004 &  1.16 & 0.362134 & 46.377322 & \imina&  \iminb     &  \iminb   & 46.739456 \\
100.0 &  0.45 & 1.142090    & 94.811929 &  1.05 & 0.464011 & 95.490008 & \imina&  \iminb     &  \iminb   & 95.954019 \\
\hline\hline
\end{tabular}
}
\end{table}

\begin{table}
\caption{Data for the first excited $^1\Sigma_u$ state: 
Total energies $E_{t1}\,E_{t2}$ and dissociation energies $E_{d1}\,E_{d2}$ 
at the corresponding equilibrium internuclear distance, 
the equilibrium internuclear distances $R_{eq1}\,R_{eq2}$,
the positions $R_{max}$, and the total energy  $E_{max}$ at the maximum,
and the total energy in the dissociation limit 
$\lim\limits_{R\to\infty} E_{t}$ as a function of the field strength $0 \leq B \leq 100$ 
(all quantities are given in atomic units).}
\begin{tabular}{dddddddddd}
\hline\hline
\multicolumn{1}{c}{\rule[-5mm]{0mm}{11mm}{\raisebox{-0.5ex}[0.5ex]{B}}} & 
\multicolumn{1}{c}{\raisebox{-0.5ex}[0.5ex]{\hspace*{0.2cm} $R_{eq1}$}} &
\multicolumn{1}{c}{\raisebox{-0.5ex}[0.5ex]{\hspace*{0.5cm} $E_{d1}$}} &
\multicolumn{1}{c}{\raisebox{-0.5ex}[0.5ex]{\hspace*{0.5cm} $E_{t1}$}} &
\multicolumn{1}{c}{\raisebox{-0.5ex}[0.5ex]{\hspace*{0.0cm} $R_{eq2}$}} &
\multicolumn{1}{c}{\raisebox{-0.5ex}[0.5ex]{\hspace*{0.5cm} $E_{d2}$}} &
\multicolumn{1}{c}{\raisebox{-0.5ex}[0.5ex]{\hspace*{0.5cm} $E_{t2}$}} &
\multicolumn{1}{c}{\raisebox{-0.5ex}[0.5ex]{\hspace*{0.0cm} $R_{max}$}} &
\multicolumn{1}{c}{\raisebox{-0.5ex}[0.5ex]{\hspace*{0.5cm} $E_{max}$}} &
\multicolumn{1}{c}{\hspace*{0.5cm} $\lim\limits_{R\to\infty} E_{tot}$} \\ \hline
0.0   &  2.09 & 0.040772 & -0.665771 &  \ia  &   \ib       &  \ib      &  \ia  &  \ib      & -0.624999 \\
0.001 &  2.08 & 0.040770 & -0.665766 &  \ia  &   \ib       &  \ib      &  \ia  &  \ib      & -0.624996 \\
0.005 &  2.09 & 0.040697 & -0.665603 &  \ia  &   \ib       &  \ib      &  \ia  &  \ib      & -0.624906 \\
0.01  &  2.09 & 0.040477 & -0.665103 &  \ia  &   \ib       &  \ib      &  \ia  &  \ib      & -0.624626 \\
0.05  &  2.12 & 0.037068 & -0.653461 &  \ia  &   \ib       &  \ib      &  \ia  &  \ib      & -0.616393 \\
0.1   &  2.13 & 0.036237 & -0.631852 &  9.37 & 0.008169 & -0.603784 &  5.16 & -0.602345 & -0.595615 \\
0.2   &  2.12 & 0.043756 & -0.583123 & 10.09 & 0.027637 & -0.567004 &  4.60 & -0.555876 & -0.539367 \\
0.5   &  2.03 & 0.065201 & -0.413216 & 10.34 & 0.068386 & -0.416401 &  4.01 & -0.382377 & -0.348015 \\
1.0   &  1.85 & 0.079251 & -0.080736 & 10.55 & 0.087513 & -0.088998 &  3.66 & -0.038398 & -0.001485 \\
2.0   &  1.59 & 0.085790 &  0.674448 & 11.46 & 0.080259 &  0.679979 &  3.34 &  0.744375 &  0.760238 \\
5.0   &  1.22 & 0.121736 &  3.201985 & 10.49 & 0.052138 &  3.271583 &  2.95 &  3.349758 &  3.323721 \\
10.0  &  0.98 & 0.178133 &  7.702800 &  9.70 & 0.011729 &  7.869204 &  2.69 &  7.955928 &  7.880933 \\
20.0  &  0.78 & 0.321481 & 17.049749 &  9.27 & 0.000271 & 17.370959 &  2.46 & 17.464313 & 17.371230 \\
50.0  &  0.57 & 0.646721 & 45.889737 &  9.04 & 0.000168 & 46.536290 &  2.21 & 46.635075 & 46.536458 \\
100.0 &  0.45 & 1.012902 & 94.733681 &  9.04 & \iminb & 95.746735 &  2.05 & 95.846945 & 95.746583 \\
\hline\hline
\end{tabular}
\end{table}

\begin{table}
\caption{Data for the second excited $^1\Sigma_u$ state: 
Total energies $E_{t1}\,E_{t2}$ and dissociation energies $E_{d1}\,E_{d2}$ 
at the corresponding equilibrium internuclear distance, 
the equilibrium internuclear distances $R_{eq1}\,R_{eq2}$,
the positions $R_{max}$, and the total energy  $E_{max}$ at the maximum,
and the total energy in the dissociation limit 
$\lim\limits_{R\to\infty} E_{t}$ as a function of the field strength $0 \leq B \leq 100$ 
(all quantities are given in atomic units).}
\begin{tabular}{dddddddddd}
\hline\hline
\multicolumn{1}{c}{\rule[-5mm]{0mm}{11mm}{\raisebox{-0.5ex}[0.5ex]{B}}} & 
\multicolumn{1}{c}{\raisebox{-0.5ex}[0.5ex]{\hspace*{0.2cm} $R_{eq1}$}} &
\multicolumn{1}{c}{\raisebox{-0.5ex}[0.5ex]{\hspace*{0.5cm} $E_{d1}$}} &
\multicolumn{1}{c}{\raisebox{-0.5ex}[0.5ex]{\hspace*{0.5cm} $E_{t1}$}} &
\multicolumn{1}{c}{\raisebox{-0.5ex}[0.5ex]{\hspace*{0.0cm} $R_{eq2}$}} &
\multicolumn{1}{c}{\raisebox{-0.5ex}[0.5ex]{\hspace*{0.5cm} $E_{d2}$}} &
\multicolumn{1}{c}{\raisebox{-0.5ex}[0.5ex]{\hspace*{0.5cm} $E_{t2}$}} &
\multicolumn{1}{c}{\raisebox{-0.5ex}[0.5ex]{\hspace*{0.1cm} $R_{max}$}} &
\multicolumn{1}{c}{\raisebox{-0.5ex}[0.5ex]{\hspace*{0.5cm} $E_{max}$}} &
\multicolumn{1}{c}{\hspace*{0.5cm} $\lim\limits_{R\to\infty} E_{tot}$} \\ \hline
0.0   &  2.03 & 0.081441 & -0.636996 & 11.12 & 0.049964 & -0.605519 &  5.65 & -0.567698 & -0.555555 \\
0.001 &  2.03 & 0.081426 & -0.636973 & 11.12 & 0.049971 & -0.605518 &  5.64 & -0.567692 & -0.555547 \\
0.005 &  2.03 & 0.081009 & -0.636427 & 11.12 & 0.050053 & -0.605471 &  5.64 & -0.567528 & -0.555418 \\
0.01  &  2.03 & 0.080107 & -0.635125 & 11.18 & 0.050291 & -0.605309 &  5.67 & -0.567116 & -0.555018 \\
0.05  &  2.04 & 0.074463 & -0.619321 & 11.32 & 0.055802 & -0.600660 &  5.45 & -0.555308 & -0.544858 \\
0.1   &  2.05 & 0.071675 & -0.594135 & 12.01 & 0.064313 & -0.586773 &  5.40 & -0.530941 & -0.522460 \\
0.2   &  2.03 & 0.062081 & -0.540641 & 15.62 & 0.059632 & -0.538198 &  5.42 & -0.477396 & -0.478566 \\
0.5   &  1.94 & 0.035148 & -0.363027 & 50.06 & 0.020138 & -0.348017 &  5.20 & -0.297160 & -0.327879 \\
1.0   &  1.78 & 0.033120 & -0.024754 &       &          &           &  4.92 &  0.058491 &  0.008366 \\
2.0   &  1.54 & 0.068597 &  0.735247 & \ia   &   \ib    &   \ib     &  4.56 &  0.858143 &  0.803844 \\
5.0   &  1.20 & 0.158473 &  3.267384 & \imina& \iminb   &  \iminb   &  4.06 &  3.490593 &  3.425857 \\
10.0  &  0.96 & 0.272912 &  7.770342 & \imina& \iminb   &  \iminb   &  3.70 &  8.118636 &  8.043254 \\
20.0  &  0.77 & 0.442635 & 17.118131 & \imina& \iminb   &  \iminb   &  3.38 & 17.648590 & 17.560766 \\
50.0  &  0.56 & 0.782055 & 45.957401 & \imina& \iminb   &  \iminb   &  3.02 & 46.845692 & 46.739456 \\
100.0 &  0.45 & 2.154264 & 94.799755 & \imina& \iminb   &  \iminb   &  2.81 & 96.074966 & 96.954019 \\
\hline\hline
\end{tabular}
\end{table}

\begin{table}
\squeezetable
\caption{Data for the third excited $^1\Sigma_u$ state: 
Total energies $E_{t1}\,-\,E_{t3}$ and dissociation energies $E_{d1}\,-\,E_{d3}$ 
at the corresponding equilibrium internuclear distance, 
the equilibrium internuclear distances $R_{eq1}\,-\,R_{eq3}$ 
and the total energy in the dissociation limit 
$\lim\limits_{R\to\infty} E_{t}$ as a function of the field strength $0 \leq B \leq 100$ 
(all quantities are given in atomic units).}
{\small
\begin{tabular}{ddddddddddd}
\hline\hline
\multicolumn{1}{c}{\rule[-5mm]{0mm}{11mm}{\raisebox{-0.5ex}[0.5ex]{B}}} & 
\multicolumn{1}{c}{\raisebox{-0.5ex}[0.5ex]{\hspace*{0.5cm} $R_{eq1}$}} &
\multicolumn{1}{c}{\raisebox{-0.5ex}[0.5ex]{\hspace*{0.5cm} $E_{d1}$}} &
\multicolumn{1}{c}{\raisebox{-0.5ex}[0.5ex]{\hspace*{0.5cm} $E_{t1}$}} &
\multicolumn{1}{c}{\raisebox{-0.5ex}[0.5ex]{\hspace*{0.1cm} $R_{eq2}$}} &
\multicolumn{1}{c}{\raisebox{-0.5ex}[0.5ex]{\hspace*{0.5cm} $E_{d2}$}} &
\multicolumn{1}{c}{\raisebox{-0.5ex}[0.5ex]{\hspace*{0.5cm} $E_{t2}$}} &
\multicolumn{1}{c}{\raisebox{-0.5ex}[0.5ex]{\hspace*{0.1cm} $R_{eq3}$}} &
\multicolumn{1}{c}{\raisebox{-0.5ex}[0.5ex]{\hspace*{0.5cm} $E_{d3}$}} &
\multicolumn{1}{c}{\raisebox{-0.5ex}[0.5ex]{\hspace*{0.5cm} $E_{t3}$}} &
\multicolumn{1}{c}{\hspace*{0.5cm} $\lim\limits_{R\to\infty} E_{tot}$} \\ \hline
0.0   & 2.00  & 0.078543 & -0.634098 &  5.67 & 0.010196 & -0.565751 &  \ia  &   \ib    &   \ib     & -0.555555 \\
0.001 & 2.00  & 0.078531 & -0.634077 &  5.67 & 0.010194 & -0.565740 &  \ia  &   \ib    &   \ib     & -0.555546 \\
0.005 & 2.00  & 0.078211 & -0.633537 &  5.67 & 0.010161 & -0.565487 & 33.89 & 0.000058 & -0.555384 & -0.555326 \\
0.01  & 2.00  & 0.077103 & -0.631765 &  5.67 & 0.010059 & -0.564721 & 34.18 & 0.000338 & -0.555000 & -0.554662 \\
0.05  & 2.01  & 0.063148 & -0.603904 &  5.56 & 0.009249 & -0.550005 & 42.09 & 0.004106 & -0.544862 & -0.540756 \\
0.1   & 2.02  & 0.060038 & -0.577452 &  5.50 & 0.009923 & -0.527337 & 36.87 & 0.005048 & -0.522462 & -0.517414 \\
0.2   & 2.00  & 0.057601 & -0.522906 &  5.79 & 0.002238 & -0.467543 & 33.92 & 0.000006 & -0.465309 & -0.465305 \\
0.5   & 1.92  & 0.062864 & -0.343463 & \ia   &  \ib        &  \ib      & 34.57 & 0.000001 & -0.280600 & -0.280599 \\
1.0   & 1.76  & 0.082384 & -0.003776 & \ia   &  \ib        &   \ib     &  \ia  & \ib      &  \ib      &  0.078608 \\
2.0   & 1.53  & 0.123535 &  0.757397 & \ia   &  \ib        &   \ib     &  \ia  & \ib      &  \ib      &  0.880932 \\
5.0   & 1.19  & 0.224179 &  3.290662 & \imina&  \iminb     &  \iminb   & \imina&  \iminb  &  \iminb   &  3.514841 \\
10.0  & 0.96  & 0.348221 &  7.794138 & \imina&  \iminb     &  \iminb   & \imina&  \iminb  &  \iminb   &  8.142359 \\
20.0  & 0.76  & 0.528386 & 17.142170 & 3.52  &  \ib & 17.707128 & \imina&  \iminb  &  \iminb   & 17.670556 \\
50.0  & 0.56  & 0.882660 & 45.981219 & 3.09  &  \ib & 46.897984 & \imina& \iminb   &  \iminb   & 46.863879 \\
100.0 & 0.45  & 1.266401 & 94.823274 & 2.82  &  \ib & 96.127449 & \imina& \iminb   &  \iminb   & 96.089675 \\
\hline\hline
\end{tabular}
}
\end{table}

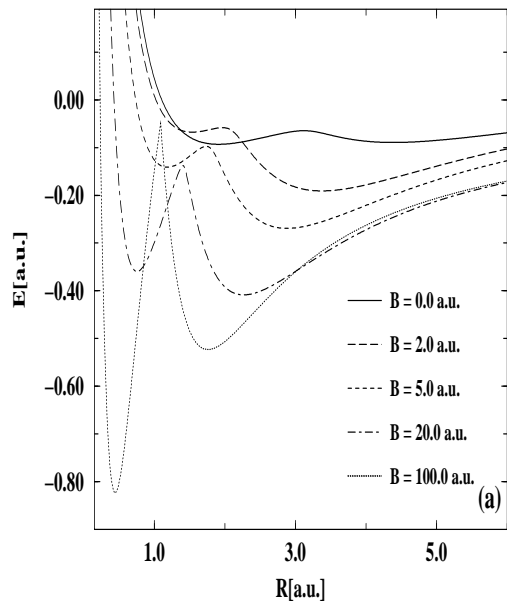
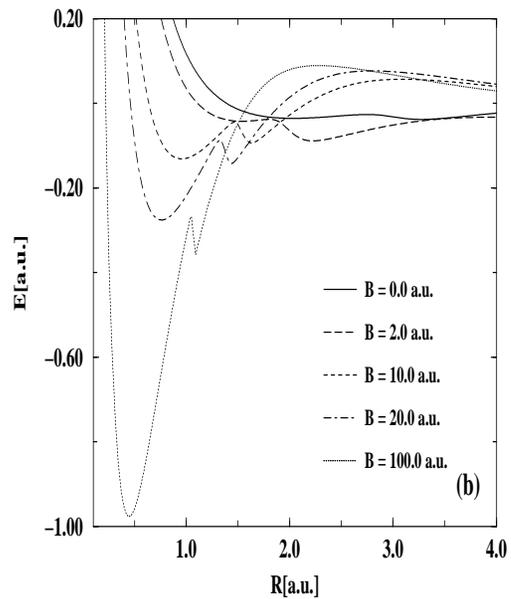
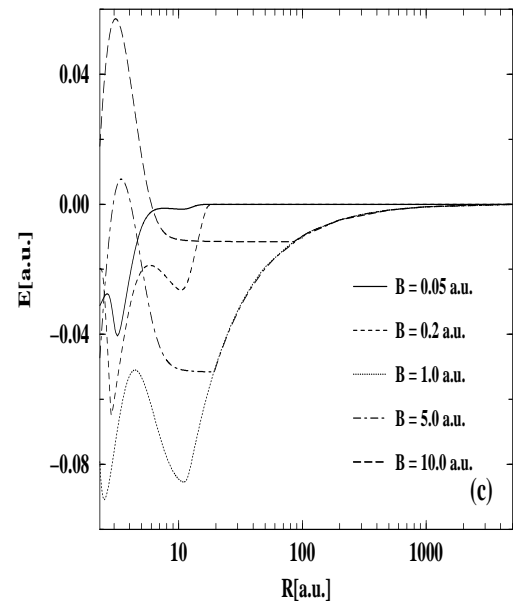
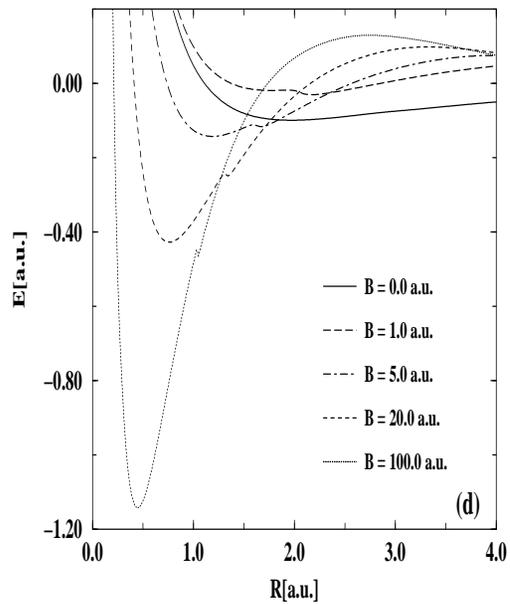
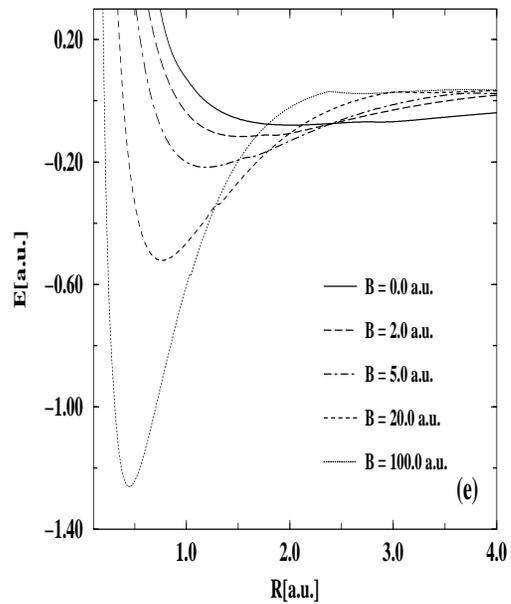
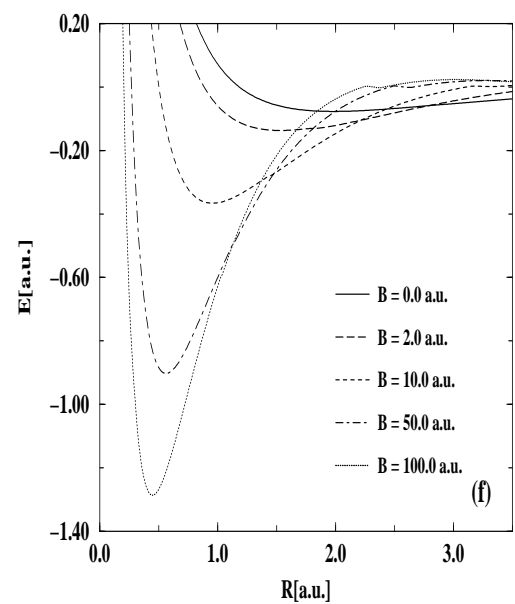
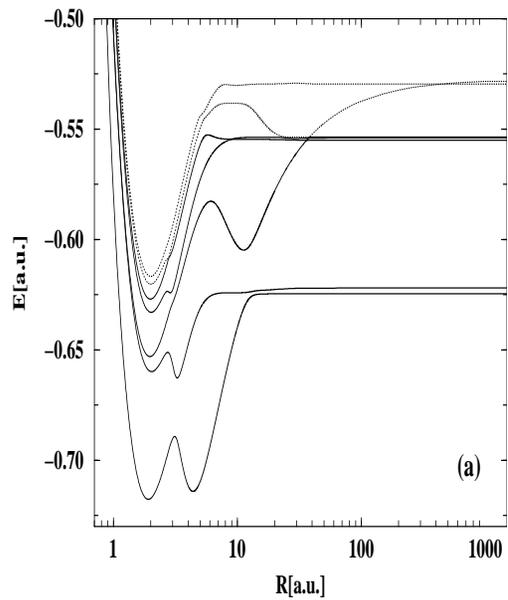
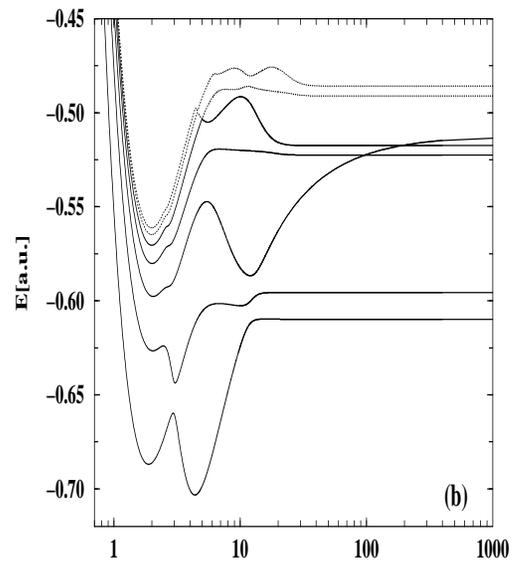
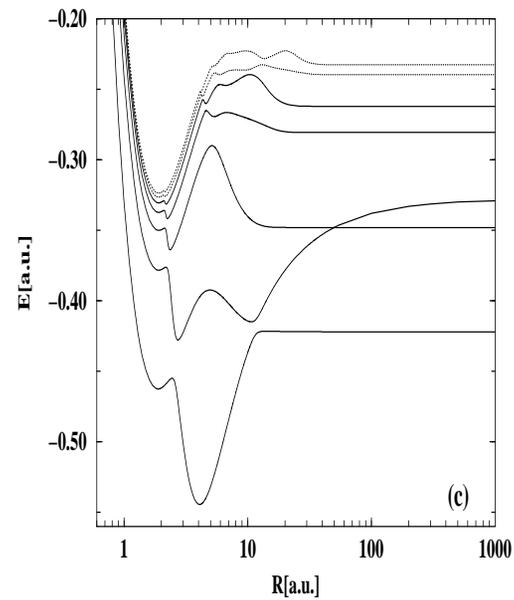
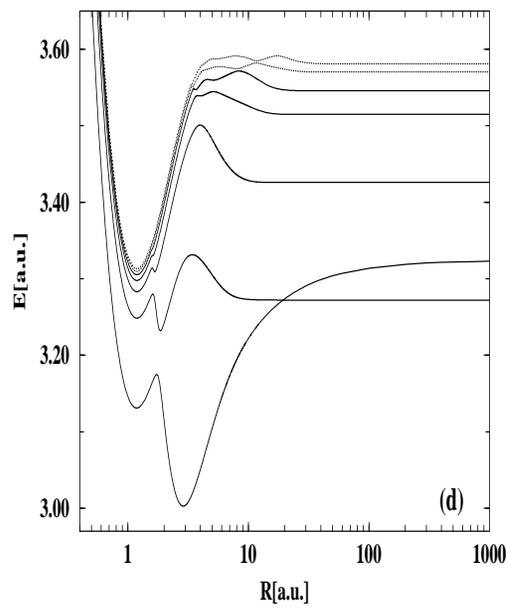
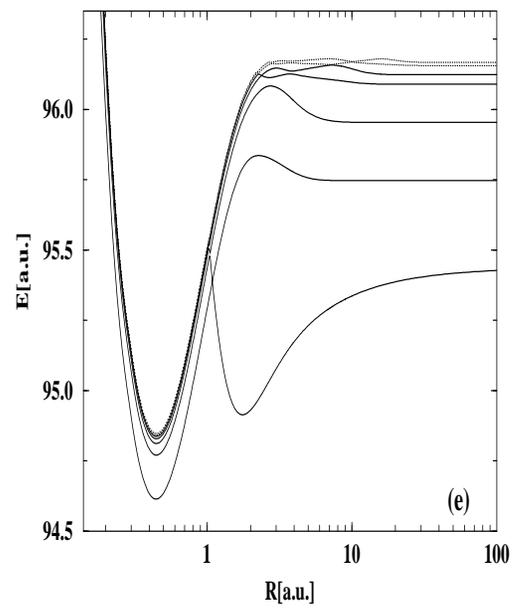
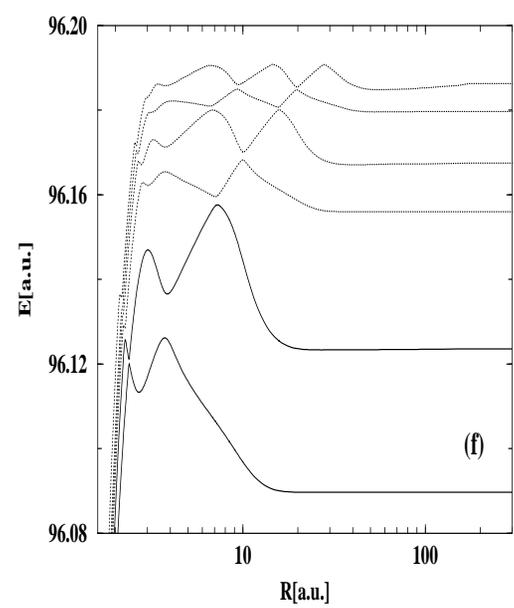
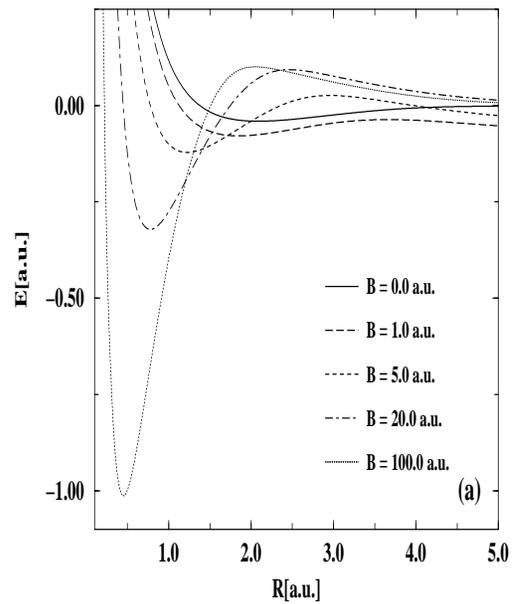
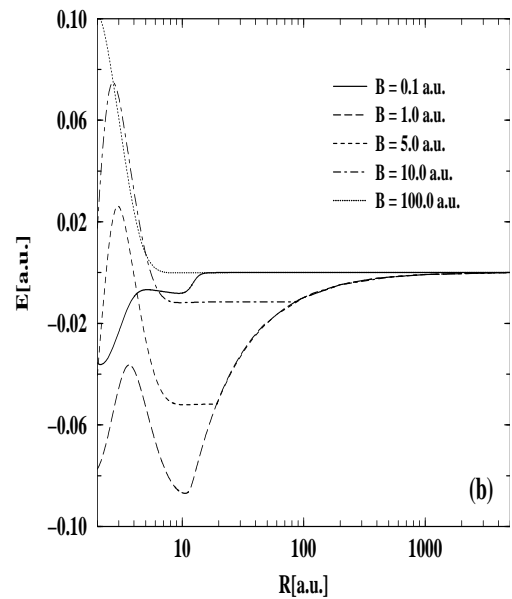
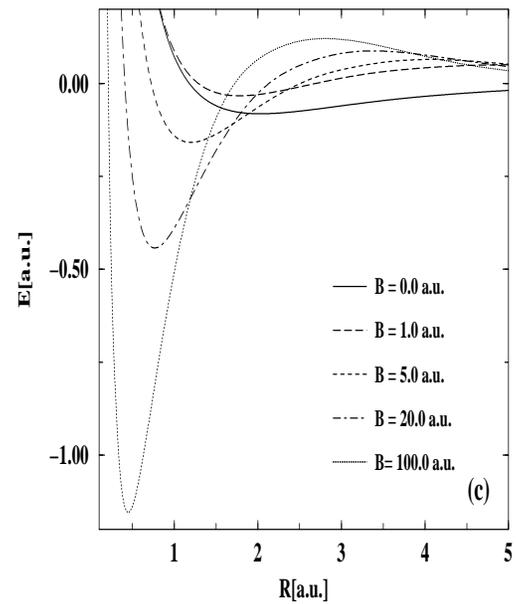
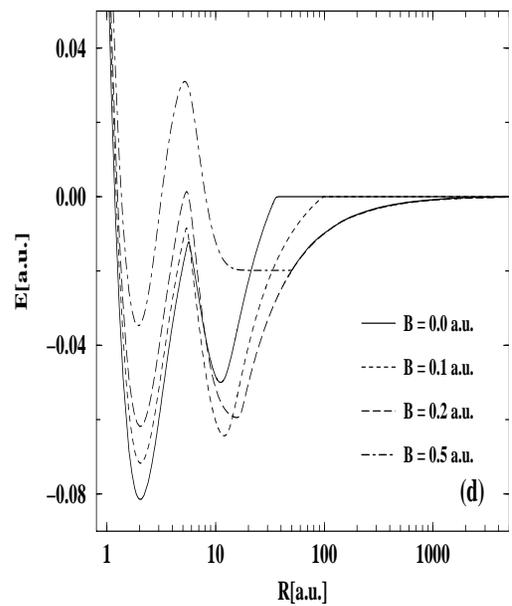
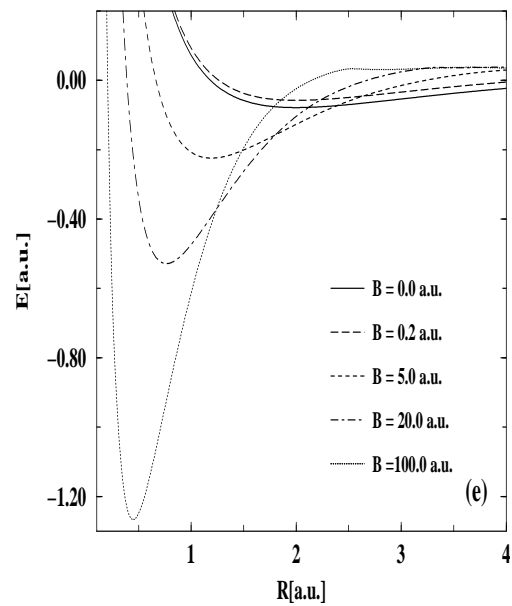
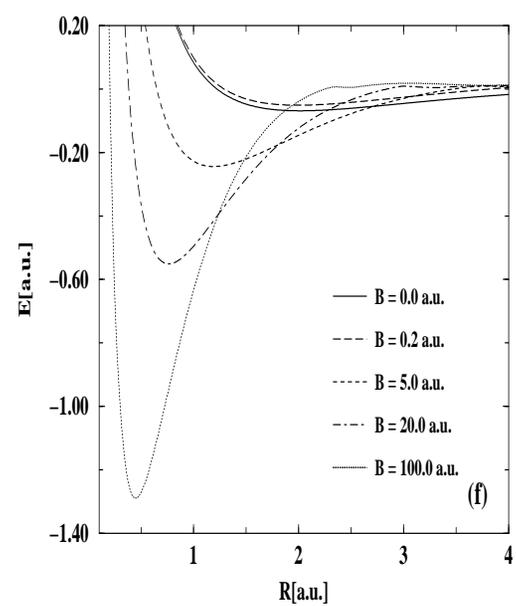
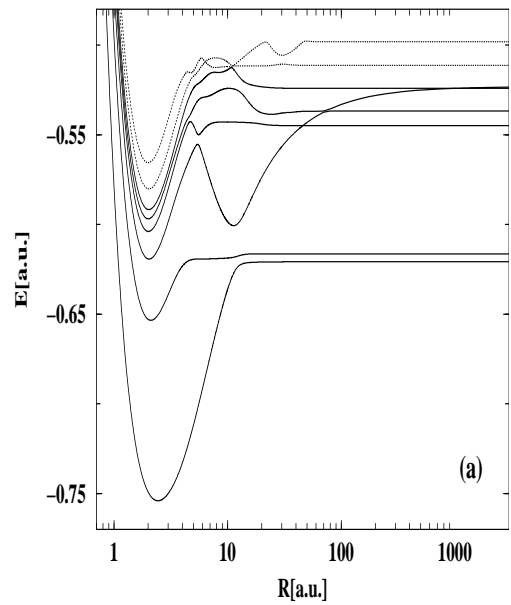
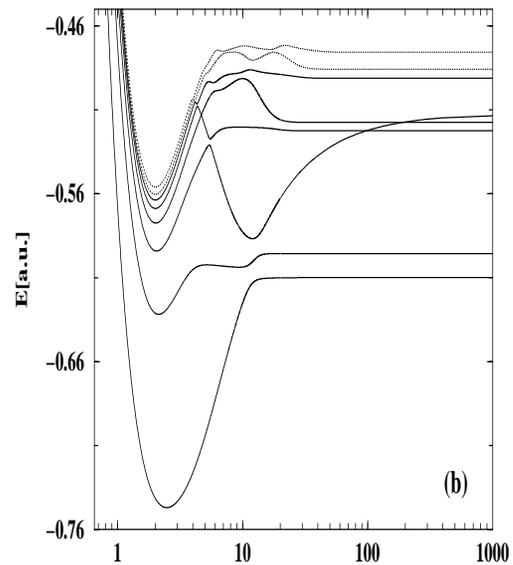
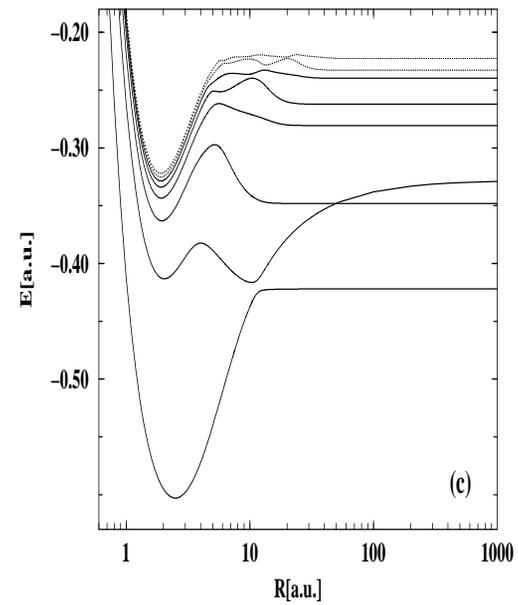
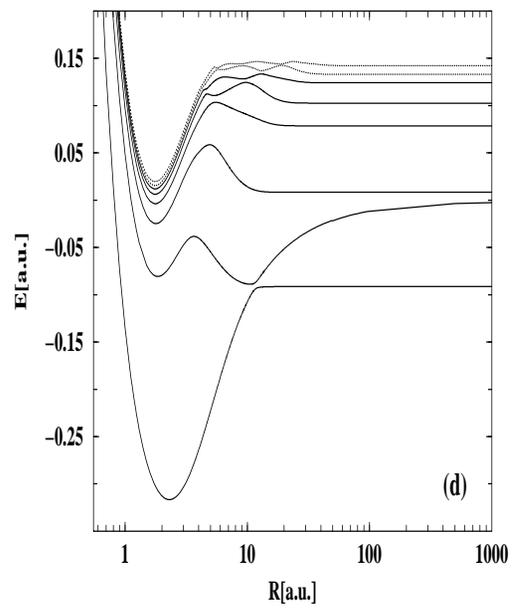
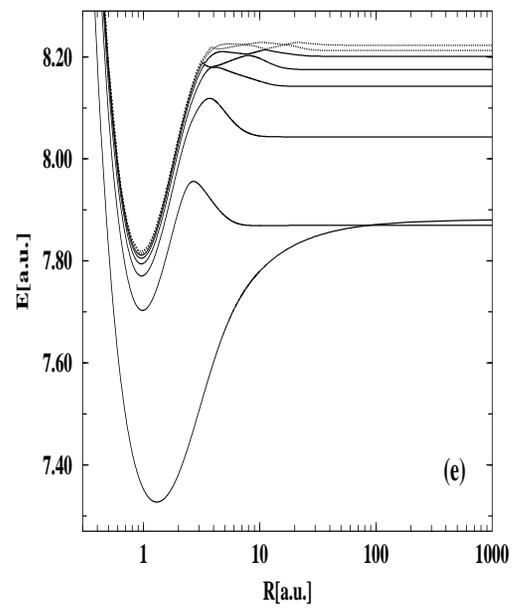
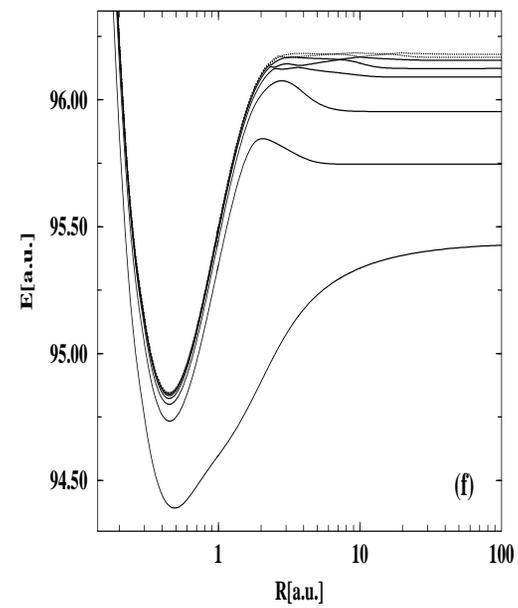
\end{document}